\documentclass[11pt,a4paper]{article}
\usepackage{graphicx}
\usepackage{subfigure}

\begin{document}

\title{\huge{A Study of Three Dimensional Edge and Corner Problems using the
neBEM Solver}}
\author{Supratik Mukhopadhyay, Nayana Majumdar}
\date{INO Section, Saha Institute of Nuclear Physics\\
      1/AF, Sector 1, Bidhannagar, Kolkata 700064, WB, India\\
      supratik.mukhopadhyay@saha.ac.in, nayana.majumdar@saha.ac.in}

\maketitle

\begin{abstract}
The previously reported \textit{neBEM} solver has been used to solve electrostatic
problems having three-dimensional edges and corners in the physical domain.
Both rectangular and triangular elements have been used to discretize the
geometries under study.
In order to maintain very high level of precision, a library of C
functions yielding exact values of potential and flux influences due to
uniform surface distribution of singularities on flat triangular and
rectangular elements has been developed and used.
Here we present the exact expressions proposed for computing the influence of
uniform singularity distributions on triangular elements and illustrate their
accuracy.
We then consider several problems of electrostatics containing edges and
singularities of various orders including plates and cubes, and L-shaped
conductors.
We have tried to show that using the approach proposed in the earlier paper on
\textit{neBEM} and its present enhanced (through the inclusion of triangular
elements) form, it is possible to obtain accurate estimates of integral features
such as the capacitance of a given conductor and detailed ones such as the
charge density distribution at the edges / corners without taking resort to
any new or special formulation.
Results obtained using \textit{neBEM} have been compared extensively with both
existing analytical and numerical results.
The comparisons illustrate the accuracy, flexibility and robustness of the new
approach quite comprehensively.
\end{abstract}


\section{Introduction}
One of the elegant methods for solving the Laplace / Poisson equations
(normally an integral expression of the inverse square law) is to set up the
Boundary Integral Equations (BIE) which lead to the moderately popular Boundary
Element Method (BEM).
In the forward collocation version of the BEM, surfaces of a given
geometry are replaced by a distribution of point singularities such as source /
dipole of unknown strengths.
The strengths of these singularities are obtained 
through the satisfaction of a given set of boundary conditions that can be
Dirichlet, Neumann or of the Robin type.
The numerical implementation requires considerable care \cite{Nagarajan93} because it
involves evaluation of singular (weak, strong and hyper) integrals.
Some of the notable two-dimensional (while all the devices are 3D by definition,
useful insight is often obtained by performing a 2D analysis) and three-dimensional
approaches used to evaluate the singular integrals are discussed in
\cite{Nagarajan93,WangTsay} and \cite{Cruse69,Kutt75,Lachat76,Srivastava92,Carini2002}
and the references in these papers.
It is well-understood that many of the difficulties in the
available BEM solvers stem from the assumption of nodal concentration of
singularities which leads to various mathematical difficulties and to the
infamous numerical boundary layers \cite{Chyuan2004,Sladek91,Ong2005} when the
source is placed very close to the field point (\cite{WangTsay} and references
[4-6] therein).
While mathematical singularities (that occur when the source and field points
coincide) have been shown to be artifacts, several techniques have been used to
remove difficulties related to physical or geometrical singularities (that occur
when boundaries are degenerate, i.e., geometrically singular, or due to a jump in
boundary conditions) such as gaussian quadrature integration, mapping techniques
for regularization, bicubic transformation, nonlinear transformation and dual BEM
techniques \cite{Chyuan2004}.
The last technique seems to be a popular one and capable of dealing with a relatively
wide range of similar problems.

Departing from the approaches mentioned in the above references and many more
to be mentioned below,
we had shown in an earlier paper \cite{EABE2006} that many of these problems
can be eliminated or reduced if we adopt a new paradigm in which the elements
are endowed with singularities distributed on them, rather than assuming the
singularities to be concentrated at specific nodal points.
Despite a large body of literature, closed form analytic
expressions for computing the effects of distributed singularities are rare
\cite{Hess67,Newman86}, complicated to implement and, often, valid only for
special cases \cite{Goto92,Lazic2006,Lazic2008}.
For example, in \cite{Hess67}, the integration of the Green function to
compute the influence of a constant source distribution is modified to an
``n-plane" integration.
The evaluation of this integration involves co-ordinate transformations and
the resulting expressions are rather complicated.
In \cite{Newman86}, the Gauss-Bonnet concept is used in which the panel is
projected onto a unit sphere and the solid angle is determined from the sum of
the induced angles.
The procedure and the resulting expressions are neither simple, nor easy to
implement in a computer code.
In fact, possibly due to these difficulties, these approaches have remained
relatively unpopular and even in very recent papers it is maintained that for
evaluating the influence due to source distributed on triangular elements in a
general case, one must apply non-analytic procedures \cite{Lazic2006}.
Thus, for solving realistic but difficult problems involving, for example,
sharp edges and corners or thin or closely spaced elements, introduction of
special formulations (usually involving fairly complicated mathematics, once again)
becomes a necessity \cite{Chyuan2004,Ong2005,Bao04}.
These drawbacks are some of the major
reasons behind the relative unpopularity of the BEM despite its significant
advantages over domain approaches such as the finite-difference and
finite-element methods (FDM and FEM) while solving non-dissipative problems
\cite{Liao2004,IEEE2006}.

The Inverse Square Law Exact Solutions (ISLES) library developed in conjunction
with the nearly exact BEM (neBEM) solver \cite{EABE2006}, in contrast, is capable of truly
modeling the effect of distributed singularities precisely and, thus,
is not limited by the proximity of other singular surfaces or their
curvature or their size and aspect ratio.
The library consists of analytic solutions for both potential and flux due to
uniform distribution of singularity on flat rectangular and triangular
elements.
These close-form exact solutions, termed as \textit{foundation expressions}, are
in the form of algebraic expressions that are
long but without complications and are fairly straight-forward to implement in
a computer program.
In deriving these foundation expressions, while the rectangular elements were
allowed to be of any arbitrary size \cite{EABE2006,NIMA2006}, the triangular
element was restricted to be a right angled triangle of arbitrary size
\cite{EMTM2NTriElem2007,arXivTriElem2007,JINST2007}.
Since any real geometry can be represented through elements of the above two
types (or by the triangular type alone), this library has allowed us to develop
the neBEM solver that is capable of solving three-dimensional potential
problems involving arbitrary geometry.
It may be noted here that any non-right-angled triangle can be easily
decomposed into two right-angled triangles.
Thus, the right-angled triangles considered here, in fact, can take
care of any three-dimensional geometry.

A set of particularly difficult problems to be dealt with by BEM is one that contains
corners and edges and, in this work, we will attempt to solve several problems
belonging to this set.
The perfectly conducting bodies studied here are unit square plate, L-shaped plate,
cube, L-shaped volume and two rectangular plates meeting at various angles and
creating an edge.
Besides being interesting and difficult, these solutions can have significant applications in
micro electromechanical systems (MEMS), nano-devices, atomic force microscopy
(AFM), electro-optical elements, micro-pattern gas detectors (MPGD) and many
other disciplines in science and technology.
For these problems, it is important to study integral features such as the
capacitance of the conductors, as well as detailed features such as the charge density,
potential and flux on various surfaces of these
objects including regions close to the geometric singularities.
While several approaches including finite-difference method (FDM),
finite element method (FEM), BEM and its variants such as the surface
charge method (SCM) and
various implementations of the Monte-Carlo technique (often coupled with Kelvin
transformation) have been used to study these problems, only the latter two
approaches, namely, BEM and Monte-Carlo technique are found to
possess the precision necessary to model the curiously difficult electrostatics with
acceptable levels of accuracy \cite{Wintle2004}.
The volume discretization methods are known to be unsuitable because of the open
nature of the problem and the inadequate representation of edge and corner singularities.
Methods using Kelvin inversion (or quadratic inversion), although accurate, have been
found incapable of handling planar problems.
It may be noted here that despite the usefulness of 2D analysis, there are an
overwhelming number of problems that need to be addressed in 3D.
As a result, several interesting approaches have been developed to analyze
edge and corner related problems in complete three dimensions, without even
the assumption of axial symmetry.
In order to maintain applicability in the most general scenarios, in this work we will
deal with the problems of edge and corner as truly three-dimensional objects
even when comparing the results with 2D analytic ones.

The problem of estimation of capacitance of square plate and cube raised to unit volt
has been studied by an especially large number of workers using entirely different
approaches.
In fact, these have been considered to be some of the major unsolved problems of
electrostatics, of which a solution is said to have been given by Dirichlet and
subsequently lost.
One of the more popular numerical approaches used to explore these problems is the
BEM / SCM \cite{Cavendish1879,Maxwell1892,Reitan57,Goto92,Read97,Read2004}.
Some of the solution attempts are  more than a century old
and yielded quite acceptable results.
The later studies \cite{Goto92,Read97,Read2004}, used the mesh refinement technique
and
extrapolation of $N$ (the number of elements used to discretize a given body) to
infinity in order to arrive at more precise estimates of the capacitance.
In order to carry out this extrapolation, uniform charge density scenario
has been maintained so that the form of charge distribution on
individual segments becomes irrelevant.
According to the \cite{Read2004}, it is justified to use uniform charge density
on individual elements because increase in complexity through the use of
non-uniform charge density ultimately does not lead to computational advantage.
In \cite{Read2004}, the author mentions that for the cube, the element
sizes are made such that the charge on each element remains approximately a
constant (independent of its distance from an edge) since this arrangement is found
to give the most accurate results.

The problem of estimation of the order of singularities at
edges and corners of different nature is strongly coupled with the problem of 
estimation of integral properties such as the capacitance of
conductors of various shapes.
Thus, this problem, which also has importance in relation to other areas of science
and technology as discussed earlier, has attracted the attention of a large number of
workers as well.
Here, fortunately, some analysis has been possible using purely theoretical
tools \cite{Jackson,Smythe}, at least for two-dimensional cases.
In \cite{Morrison76}, the authors used a
singular perturbation technique to obtain the singularity index at inside and outside corners
of a sectorial conducting plate.
Similarly, corner singularity exponents were numerically obtained in \cite{Read2004}.
According to \cite{Greenfield2004}, it was possible to achieve accuracy of
one in million through the use of FEM approximations for both electrodes and
surface charge density, in addition to proper handling of edge and corner
singularities.
In this investigation, the Fichera's theorem was used to correctly describe the
peculiarities of surface charge density behaviour in the vicinity of the electrode
ribs and tips.

According to another recent work \cite{Ong2005}, low-order polynomials used
to represent
the corners and edges lead to errors in estimation of the derivatives of the
potential, and that is the crux of the problem.
Despite its advantages (no prior knowledge of singular elements and the order of
singularity is necessary) and good convergence characteristics, the mesh refinement
approach has been mentioned to be less accurate than two other methods, namely,
1) singular elements, and
2) singular functions.
Among these, beforehand knowledge of the location and behaviour of the
singularity in terms of the order of singularity is
required for (1).
The singular function approach (2) performs the best since it uses both the
above and also information on singularity profile
(corresponding to the eigenvalues and eigenvectors of the given geometry).
Unfortunately, this approach requires complex formulation and is generally restricted to 2D.
Since the singular element approach uses the location information and only the order
of the singularity, it is more versatile and popular.
Thus, \cite{Ong2005} implements the method of singular elements in 3D while attempting
to use proper shape function /
interpolation function to model correct singular behaviour at corners and edges.
It is shown that singular elements are needed to accurately capture the
behavior at singular regions, such as sharp corners and edges, where standard
elements fail to give an accurate result.
Unfortunately, singular elements must be defined before it is possible to apply either
the singular function or the singular element approach.
This is a nontrivial task since for a realistic device, there can
easily be thousands, if not millions, of elements involved.
According to \cite{Su2002}, the manual classification of boundary elements based
on their singularity conditions is an immensely laborious task if not outright impractical.
In \cite{Su2002}, the authors developed an algorithm to extract the regions
where singularity arises by querying the geometric model for
convex edges based on geometric information of the model.
The associated nodes of the boundary elements on these edges were then retrieved
and categorized according to different types of singularity configuration.
The algorithm developed was implemented in the PATRAN command language (PCL) on
the MSC/PATRAN platform \cite{MSC} which is an
industry standard finite-element pre- and post-processor allowing a high
degree of customization.
In order to determine the order of singularity, two-dimensional results have been used
directly \cite{Jackson,Bazant}
for edges under the assumption that the point lies sufficiently far from any corner.
In fact, in \cite{Su2002}, any point other than the vertices has been
considered to be far enough.
This approach is quite unlikely to be very accurate, especially when the
dimensions of the devices are small enough so that no point may even be
considered to be far enough.
Later, while discussing the results obtained using our proposed approach, we will
return to this issue again.
A study on the effect of bias ratio (a ratio of the largest element length to the
smallest element length, similar to r- mesh refinement) has led the
authors  \cite{Ong2005} to state that while for normal elements this ratio
should be around 4:1 for a given problem, the singular element approach works
better with lower bias ratio.

Besides the above procedures to identify the singular elements and their orders,
attempts have also been made to place nodes at optimal positions while solving
problems involving edges and corners.
For example, in fracture mechanics \cite{Ariza97}, for handling singularities of
the order 0.5, mid-side nodes in quadratic elements is normally shifted to quarter
point positions.
Several other techniques such as least square, constrained displacement of side nodes
adjacent to a crack-tip \cite{Abdi89} have also been used
to find the optimum position of the mid-side nodes for handling singularities
of other orders.
In \cite{Jun98}, the authors have subsequently shown that this approach of
shifting mid-side nodes is likely to produce singularities of order 0.5 only and it is not
suitable to impose arbitrary singularity in isoparametric elements by simply shifting
side nodes to assumed positions.

In addition to BEM approaches, work has also been carried out using radically
different approaches, for example, the Monte Carlo methods.
An impressive array
of work exist providing very accurate estimates of capacitance and variation
of charge density near edges and corners
\cite{Zhou94,Given97,Hwang2001,Mansfield2001,Mascagni2004,Wintle2004,Hwang2005}.
According to \cite{Wintle2004}, the use of BEM introduces unnatural estimate of the
charge density distribution - not for shapes with smooth contours (discs or spheres),
but for plates and cubes.
It has been mentioned that the situation is worst for the corner singularities
of plates in which case there are no other surfaces present (as in a cube) to
weaken the order of singularity at the corner.
We will discuss this issue when results for square plates are presented below.
The other point is that the BEM can not satisfy the boundary conditions at or,
at least, close to the edge
because, the collocation points in BEM does not match the boundary of the device.
In fact, it is mentioned that the
solutions become unstable when the collocation points are shifted away from the
centroid of the elements.
This notion, as mentioned in the earlier paragraph, is not without counter-examples.
Moreover, this is a point that we plan to take up in a future communication where we
hope to illustrate that for the proposed formulation, shifting the collocation points to
non-centroid locations does not lead to numerical instabilities.
In addition, the approach of extrapolating capacitance has been criticized on the
ground that they do not match for different amounts of shift.
It has been mentioned that no formal error analysis exists for methods other than
FDM and the extrapolation is purely empirical in nature.
It has been observed that the apparent high accuracy may be illusory in nature and
citing \cite{Sato87}, it has been emphasized that situation can become even worse by
attempting extrapolation of results obtained using non-equivalent meshes.

By developing a model that incorporates the truly distributed nature of
sources / doublets / vortices on surfaces of three-dimensional geometries, we have
recently shown \cite{EABE2006,NIMA2006,JINST2007} that it is possible to use the
same formulation for studying a very wide range of problems (multi-scale, involving
multiple layers of dielectric materials) governed by the Poisson's equation.
Recently, we have extended the new formulation with the capability of including
triangular elements as an option for discretizing arbitrary three-dimensional bodies
\cite{EMTM2NTriElem2007,arXivTriElem2007}.
Here, we present the expressions to evaluate the exact values of potential and fluxes
at any arbitrary point due to uniform singularity distributed on right-angled
triangular element.
These expressions have been included as additional functions of the ISLES library and
subsequently used in the neBEM solver in the manner usual to, probably, most
of the BEM solvers available.
Besides presenting the expressions, we have also shown results to illustrate the
accuracy of the expressions under various circumstances.

In this report, we have presented studies on the electrostatic configuration of several
three-dimensional bodies, all of which contain corners and / or edges.
The classic benchmark problems of estimating the capacitance of a unit square
plate and unit cube raised to unit volt have been addressed.
Electrostatics of generic shapes such as L-shaped plate and L-shaped 3D conductors
has also been analyzed.
In all the above cases, the order of the singularity distribution near the edges and corners
has been estimated in addition to estimating the capacitance of the conductors
themselves.
The singularity distributions obtained have been compared with theoretical and numerical
studies carried out by earlier workers.
The variation of the singularity along an edge between two corners at its
ends has been studied, probably for the first time.
Finally, the well-known problem of electrostatic configuration of two planes
intersecting at different angles has been addressed.
The two-dimensional counter-part of this problem is known to have analytic
solution and has even been discussed in several textbooks on electromagnetics
\cite{Jackson,Smythe}.
Although accessible analytically, this problem seems to have been rarely solved using
numerical techniques \cite{Integrated}.
This benchmark problem is known to be a difficult one and, in order to test the proposed
approach under difficult circumstances, we have computed electrostatic properties
of a three-dimensional analogue of this problem close to the point of intersection for a
very wide range of angle of intersection.
Following the above studies, we have come to the conclusion that the proposed approach
is capable of solving critical multi-scale problems governed by the Poisson's equation
in a rather straight-forward manner.
While higher bit accuracy, improved evaluation of transcendental functions, adaptive mesh
generation and parallelization is expected to be of significant help,
no special mathematical treatment or new formulation has been found to be necessary to
deal with problems involving corners / edges and extremely closely spaced surfaces.

According to \cite{Wintle2004}, using BEM, it is difficult to obtain physically
consistent results close to these geometric singularities.
Wild variations in the magnitude of the charge density have been
observed with the change in the degree of discretization, the reason once
again being associated with the nodal model of singularities.
In contrast, using neBEM, we have obtained very smooth variation close to corner.
Presence of oscillations seemingly acceptable to \cite{Wintle2004}, has not occurred.
In fact, oscillations close to edges and corners considered in this work seem to
indicate numerical inaccuracy and have been treated accordingly.
In addition to the shape, the magnitudes of the charge density have been
found to be consistently converging to physically realistic values.
These results clearly indicate that since the foundation expressions of the solver
are exact, it is possible to find the potential and flux accurately in
the complete physical domain, including the critical near-field domain using
neBEM.
In addition, since the singularities are no longer assumed to be nodal and
we have the exact expressions for potential and flux throughout the physical
domain, the boundary conditions no longer need to be satisfied at
\textit{special} collocation points such as the centroid of an element.
Although consequences of this considerable advantage is still under study, it is
expected that this feature will allow neBEM to yield even more accurate estimates for
problems involving corners and edges since it should be possible to generate an
over-determined system of equations by placing extra collocation points near the edges /
corners.
This will also allow the method to satisfy the boundary conditions of a given geometry
at its true boundaries.

It should be noted here that the exact expressions for triangular elements
consist of a significantly larger number of mathematical operations than those for
rectangular elements presented in \cite{EABE2006}.
Thus, for any solver based on the ISLES library, it is more economical to use a mixed
mesh of rectangular and triangular elements using rectangular elements as much as
possible.
\section{Governing equations and Exact Solutions}
In the following discussions, we will concentrate on the electrostatics of conducting
geometries governed by the Poisson's equation.
Using BEM approach, the Poisson's equation for electrostatic potential
\[
\nabla^2 \phi(\vec{r}) = - \rho(\vec{r}) / \epsilon_0
\]
can be solved to obtain the distribution of charges which leads to a given
potential configuration.
For a point charge $q$ at $\vec{r}^{\,\prime}$ in 3D space, the potential
$\phi(\vec r)$ at $\vec{r}$ is known to be
\[
\phi(\vec{r}) = \frac {q} {4 \pi \epsilon_0 | \vec{r} - \vec{r^{\prime}} |}
\]
For a general charge distribution with charge density $\rho(\vec {r^{\prime}})$,
superposition holds and results in
\begin{equation}
\label{eq:Potential}
\phi(\vec{r}) = \int \frac{ \rho (\vec {r^{\prime}}) d{v^{\prime}} }
            { 4 \pi \epsilon_0 | \vec{r} - \vec{r^{\prime}} | }
    = \int G(\vec{r}, \vec{r^{\prime}}) \rho(\vec{r^{\prime}}) d{v^{\prime}}
\end{equation}
where
\[
G(\vec{r}, \vec{r^{\prime}})
    = \frac {1} {4 \pi \epsilon_0 | \vec{r} - \vec{r^{\prime}} |}
\]
is the free space Green's function for the Laplace operator in 3D
with $\epsilon_0$, the permittivity of free space.
Similarly, the field for a general charge distribution can be written as
\[
\vec{E}(\vec{r}) = - \nabla \phi
\]
leading to
\[
\vec{E}(\vec{r}) = - \nabla
	\left ( \int G(\vec{r}, \vec{r^{\prime}}) \rho(\vec{r^{\prime}}) d{v^{\prime}} \right )
\]
and, finally to,
\begin{equation}
\label{eq:Field}
\vec{E}(\vec{r}) 
	=  \int \frac{ \rho (\vec{r^{\prime}})(\vec{r} - \vec {r^{\prime}}) d{v^{\prime}} }
                { 4 \pi \epsilon_0 | \vec{r} - \vec{r^{\prime}} |^3 }
\end{equation}
The charge distribution can be obtained from equation (\ref{eq:Potential}) or
(\ref{eq:Field}) by satisfying the boundary conditions at collocation points
known either in the form of potential (Dirichlet)
or flux (Neumann) or a mixture of these two (Mixed/Robin) on material
boundaries/surfaces present in the domain.

Considering the Dirichlet problem only at present (for ease of discussion), the
following integral equation of the first kind can be set up.
\begin{equation}
\label{eq:Green}
\phi(\vec r)
= \int_{vol} G(\vec r, \vec {r^{\prime}}) \rho(\vec {r^{\prime}}) d {v^{\prime}}
\end{equation}
In the above equation, $\phi(\vec r)$ is the potential at a point $\vec r$ in
space and $\rho(\vec r^{\,\prime})$ is the charge density at an infinitesimally
small volume $dv^{\prime}$ placed at $\vec {r^{\prime}}$.
The problem is, generally, to find $\rho(\vec {r^{\prime}})$ as a function of space
resulting the known distribution of $\phi(\vec r)$.
Once the charge distribution on the boundaries and all the surfaces is known,
potential and field at any point in the computational domain can be 
obtained using the same equation (\ref{eq:Green}) and its derivative.

The primary step of the BEM technique is to discretize the boundaries and
surfaces of a given problem.
The elements resulting out of the discretization process are normally
rectangular or triangular though elements of other shapes are also used.
Elements of triangular shape can be used to model geometries of any variety and,
thus, is one of the most commonly used in many approaches of numerical
simulation including FEM and BEM.
In the collocation approach, the next step is to find out charge distribution on the
elements that satisfies equation (\ref{eq:Green}) following the given boundary
conditions.
The charge distribution is normally represented in terms of known basis functions with
unknown coefficients.
For example, in zero-th order formulations using constant basis function, which
is also the most popular one among all the BEM formulations because of a good
optimization between accuracy and computational complexity, the charge
distribution on each element is assumed to be uniform and equivalent to a point
charge located at the centroid of the element.
This is the method that is referred to as the {\it usual BEM} in the 
rest of the paper.
However, diverse varieties of basis function have been exercised to develop
many more BEM formulations in order to represent the charge distribution on an
element more efficiently so as to enhance the accuracy of the method.
Since the potentials on the surface elements are known from the given potential
configuration, equation (\ref{eq:Green}) can be used to generate algebraic expressions
relating unknown charge densities and potentials at the centroid of the elements.
One unique equation can be obtained for each centroid considering
influences of all other elements including self influence and, thus, 
the same number of equations can be generated as there are unknowns.
In matrix form, the resulting system of simultaneous linear algebraic set of
equations can be written as follows
\begin{equation}
\label{eq:Matrix}
\mathbf{K} \cdot \mathbf{\rho} = \mathbf{\phi}
\end{equation}
where $\mathbf{K}$ is the matrix consisting of influences among the elements due to
unit charge density on each of them, $\mathbf{\rho}$ represents a column vector of 
unknown charge densities at centroids of the elements and $\mathbf{\phi}$
represents known values of potentials at the centroids of these elements.
Each element of this influence coefficient or capacity coefficient matrix, 
$\mathbf{K}$ is a direct evaluation of an equation similar to equations 
(\ref{eq:Potential}) or (\ref{eq:Field}) which represents the effect of a
single element on a boundary/surface (obtained through discretization) on a
point where a boundary condition of the given problem is known.
While, in general, this should necessitate an integration of the Green's function
over the area of the element, this integration is avoided in most of the BEM
solvers through the assumption of nodal concentration of singularities with
known basis function.
The construction of the matrix implies that its diagonal elements are dominant
through the presence of the Coulomb-type singularity in the kernel.
This singularity has been shown to make the solutions well-defined in the
class of rather smooth functions \cite{Greenfield2004}.
Since the right hand side of (\ref{eq:Matrix}) is known, in principle, it
is possible to solve the system of algebraic equations and obtain surface
charge density on each of the element used to describe the conducting surfaces
of the detector following
\[
\mathbf{\rho} = \mathbf{K}^{-1} \cdot \mathbf{\phi}
\]
Once the charge density distribution is obtained, equations
(\ref{eq:Potential}) and (\ref{eq:Field}) can be used to obtain both 
potential and field at any point in the computational domain.

Despite the elegance of formulation, the usual BEM suffers from several
drawbacks that have resulted in its relative lack of popularity.
Two of the most important ones can be mentioned as follows.
(i) It is assumed that a surface distribution of charge density on an 
element can be represented by a nodal arrangement based on a chosen basis
function.
(ii) It is assumed that the satisfaction of the boundary condition at a
predetermined point (or, through the use of known shape functions) is 
equivalent to satisfying the same on the whole element in a distributed manner.
The former assumption leads to infamous numerical boundary layer
\cite{Chyuan2004,Sladek91,Ong2005,WangTsay}
due to which the near-field solution in regions close to an element becomes erroneous.
Thus the estimation of potential and field in near-field region close to the
boundaries and surfaces by usual BEM is found to be inaccurate.
This also leads to complications in solving problems involving closely spaced
surfaces such as degenerate surfaces, edges, corners and other geometrical 
singularities. 
The degenerate surface refers to a boundary, two portions of which approach 
each other such that the exterior region between the two portions becomes 
infinitely thin.
It is well known that the coincidence of two boundaries 
gives rise to an ill-conditioned problem.
A number of special formulations 
has been developed to cope up with these
problems but, unfortunately, most of these formulations are effective in a
rather small subset of problems related to potential and field that are usually
faced in reality.

This problem has been resolved to a great extent through the development of the
neBEM solver that uses {\it exact integration} of the Green's 
function and its derivative in its formulation.
These integrations for rectangular and triangular elements having uniform
charge density have been obtained as closed-form analytic expressions using
symbolic mathematics \cite{EABE2006}.
Thus they account for truly distributed nature of charge density on a given
element.
Besides the fundamental change in the way the influence coefficient matrix
is computed and the foundation expressions used for evaluating potential and
field at any point after the charge density vector is solved for, most of
the other features of neBEM are similar to any other BEM solver.

The expressions for potential and flux at a point $(X, Y, Z)$ in free space due
to uniform source distributed on a rectangular flat surface having corners
situated at $(x_1, z_1)$ and $(x_2, z_2)$ has been presented, validated and
used in \cite{EABE2006, NIMA2006} and, thus, is not being repeated here.

\begin{figure}[hbt]
\begin{center}
\includegraphics[height=3in,width=3in]{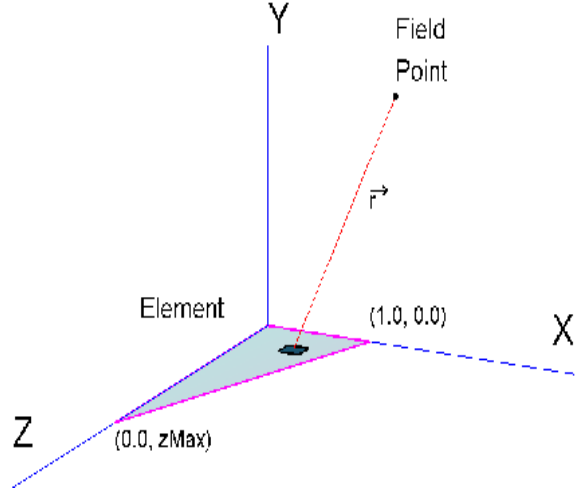}
\caption{\label{fig:TriElemGeom} Right-angled triangular element with x-length 1
and an arbitrary z-length, $z_M$; P is the point where the influence (potential
and flux) is being computed}
\end{center}
\end{figure}

Here, we present the exact expressions necessary to compute the potential and
flux due to a right-angled triangular element of arbitrary size, as shown in
Fig.\ref{fig:TriElemGeom}. It may be noted here that the length in the X
direction has been normalized, while that in the Z direction has been allowed to
be of any arbitrary magnitude, $z_M$. From the figure, it is easy to see
that in order to find out the influence due to triangular element, we have
imposed another restriction, namely, the necessity that the X and Z axes
coincide with the perpendicular sides of the right-angled triangle.
Both these restrictions are trivial and can be taken care of by carrying out
suitable scaling and appropriate vector transformations. It may be noted here
that closed-form expressions for the influence of rectangular and triangular
elements having uniform singularity distributions have been previously presented
in \cite{Newman86,Goto92}. However, in these works, the expressions presented
are quite complicated and difficult to implement. In \cite{EABE2006} and in the
present work, the expressions we have presented are lengthy, but completely
straight-forward. As a result, the implementation issues of the present
expressions, in terms of the development of the ISLES library and the neBEM
solver, are managed quite easily.
It is easy to show that the influence (potential) at a point $P(X,Y,Z)$ due to
uniform source distributed on a right-angled triangular element as
depicted in Fig.\ref{fig:TriElemGeom} can be represented as a multiple of
\begin{equation}
\Phi(X,Y,Z) = \int_{0}^{1} \int_{0}^{z(x)}
            \frac{dx\,dz}{\sqrt{(X-x)^2 + Y^2 + (Z-z)^2}}
\label{eqn:PotTriInt}
\end{equation}
in which we have assumed that $x_1=0$, $z_1=0$, $x_2=1$ and $z_2=z_M$, as shown
in the geometry of the triangular element.
The closed-form expression for the
potential has been obtained using symbolic integration \cite{MatLabBook} which
was subsequently simplified through substantial effort. It is found to be
significantly more complicated in comparison to the expression for
rectangular elements presented in \cite{EABE2006} and can be written as
\begin{eqnarray}
\label{eqn:PotTriExact}
\lefteqn{\Phi =} \nonumber \\
&& \frac{1}{2}
	\left( \right. (z_M Y^2 - X G) (LP_1 + LM_1 - LP_2 - LM_2) 
+ i \left| Y \right| (z_M X + G) (LP_1 - LM_1 - LP_2 + LM_2) \nonumber \\
&& - S_1 X ( tanh^{-1} (\frac{R_1 + i I_1}{D_{11} |Z|})
           + tanh^{-1} (\frac{R_1 - i I_1}{D_{11} |Z|}) 
         - tanh^{-1} (\frac{R_1 + i I_2}{D_{21} |Z|})
           - tanh^{-1} (\frac{R_1 - i I_2}{D_{21} |Z|}) ) \nonumber \\
&& + i S_1 |Y| ( tanh^{-1} (\frac{R_1 + i I_1}{D_{11} |Z|})
               - tanh^{-1} (\frac{R_1 - i I_1}{D_{11} |Z|}) 
             - tanh^{-1} (\frac{R_1 + i I_2}{D_{21} |Z|})
               + tanh^{-1} (\frac{R_1 - i I_2}{D_{21} |Z|}) ) \nonumber \\
&& + \frac{2 G}{\sqrt{1+{z_M}^2}}
\log \left( \right. \frac{\sqrt{1+{z_M}^2} D_{12} - E_1}
				{\sqrt{1+{z_M}^2} D_{21} - E_2} \left. \right) 
 + 2 Z \log \left( \frac{D_{21} - X + 1}{D_{11} - X} \right) \left. \right) + C
\end{eqnarray}

where, 
\[
D_{11} = \sqrt { (X-x_1)^2 + Y^2 + (Z-z_1)^2 }; \,
D_{12} = \sqrt { (X-x_1)^2 + Y^2 + (Z-z_2)^2 }
\]
\[
D_{21} = \sqrt { (X-x_2)^2 + Y^2 + (Z-z_1)^2 }; \,
I_1 = (X-x_1)\,\left| Y \right|; \, I_2 = (X-x_2)\,\left| Y \right|
\]
\[
S_1 = {\it sign} (z_1-Z);\, R_1 = Y^2 + (Z-z_1)^2
\]
\[
E_1 = (X + {z_M}^2 - z_M Z);\, E_2 = (X - 1 - z_M Z), \,
\]
\[
G = z_M (X - 1) + Z; \,
H_1 = Y^2 + G (Z - z_M);\, H_2 = Y^2 + G Z
\]
\begin{eqnarray*}
LP_1 =
	 \frac{1}{G - i z_M |Y|} 
	log(\frac{(H_1\, + G D_{12}) + i |Y| (E_1 - i z_M D_{12})} {- X + i |Y|})
\end{eqnarray*}
\begin{eqnarray*}
LM_1 = 
   \frac{1}{G + i z_M |Y|} 
	log(\frac{(H_1\, + G D_{12}) - i |Y| (E_1 - i z_M D_{12})} {- X - i |Y|})
\end{eqnarray*}
\begin{eqnarray*}
LP_2 = 
   \frac{1}{G - i z_M |Y|} 
	log(\frac{(H_2\, + G D_{21}) + i |Y| (E_2 - i z_M D_{21})} {1 - X + i |Y|})
\end{eqnarray*}
\begin{eqnarray*}
LM_2 = 
   \frac{1}{G + i z_M |Y|} 
	log(\frac{(H_2\, + G D_{21}) - i |Y| (E_2 - i z_M D_{21})} {1 - X - i |Y|})
\end{eqnarray*}
and $C$ denotes a constant of integration.

Similarly, the flux components due to the above singularity distribution can
also be represented through closed-form expressions as shown below:

\begin{eqnarray}
\label{eqn:FxTriExact}
\lefteqn{F_x = -\frac{\partial \Phi}{\partial x} =} \nonumber \\
&& \frac{1}{2}
	\left( \right. (G) (LP_1 + LM_1 - LP_2 - LM_2) 
 - i \left| Y \right| (z_M) (LP_1 - LM_1 - LP_2 + LM_2) \nonumber \\
&& + S_1 ( tanh^{-1} (\frac{R_1 + i I_1}{D_{11} |Z|})
           + tanh^{-1} (\frac{R_1 - i I_1}{D_{11} |Z|}) 
         - tanh^{-1} (\frac{R_1 + i I_2}{D_{21} |Z|})
           - tanh^{-1} (\frac{R_1 - i I_2}{D_{21} |Z|}) ) \nonumber \\
&& + \frac{2 z_M}{\sqrt{1+{z_M}^2}}
\log \left( \right. \frac{\sqrt{1+{z_M}^2} D_{12} - E_1}
				{\sqrt{1+{z_M}^2} D_{21} - E_2} \left. \right) \left. \right)
				+ C
\end{eqnarray}

\begin{eqnarray}
\label{eqn:FyTriExact}
\lefteqn{F_y = -\frac{\partial \Phi}{\partial y} =} \nonumber \\
&& \frac{-1}{2}
	\left( \right. (2 z_M Y) (LP_1 + LM_1 - LP_2 - LM_2) 
 + i \left| Y \right| (Sn(Y) G) (LP_1 - LM_1 - LP_2 + LM_2) \nonumber \\
&& + i S_1 Sn(Y) ( tanh^{-1} (\frac{R_1 + i I_1}{D_{11} |Z|})
               - tanh^{-1} (\frac{R_1 - i I_1}{D_{11} |Z|}) 
             - tanh^{-1} (\frac{R_1 + i I_2}{D_{21} |Z|})
         + tanh^{-1} (\frac{R_1 - i I_2}{D_{21} |Z|}) ) \left. \right) 
			+ C \nonumber \\
&&
\end{eqnarray}

and,

\begin{eqnarray}
\label{eqn:FzTriExact}
F_z = -\frac{\partial \Phi}{\partial z} =
\left( \right . \frac{1}{\sqrt{1+{z_M}^2}}
\log \left( \right. \frac{\sqrt{1+{z_M}^2} D_{21} - E_2}
				{\sqrt{1+{z_M}^2} D_{12} - E_1} \left. \right) 
 + \log \left( \frac{D_{11} - X}{D_{21} - X + 1} \right) \left. \right) + C
\end{eqnarray}
where $Sn(Y)$ implies the sign of the Y-coordinate and $C$ indicates constants
of integrations. It is to be noted that the constants of different integrations
are not the same.
In addition to being extremely useful in the mathematical modeling of physical
processes governed by the inverse square laws, these expression are expected
to be useful as benchmark expressions for other approximate formulations.
Being exact and valid throughout the physical domain, they can be used to
formulate versatile solvers to solve multi-scale multi-physics problems governed
by the Laplace / Poisson equations involving Dirichlet, Neumann or Robin
boundary conditions.

\section{Results and Discussions}
\subsection{Exact expressions}

\begin{figure}[hbt]
\begin{center}
\includegraphics[height=4in,width=4in]{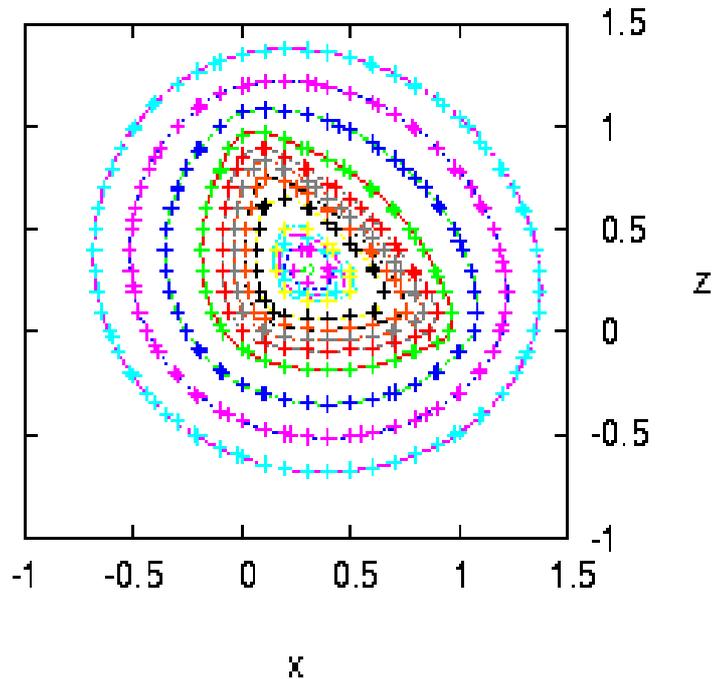}
\caption{\label{fig:CntrTriElem} Potential contours on a triangular element computed using exact expressions and by numerical quadrature}
\end{center}
\end{figure}

\begin{figure}[hbt]
\begin{center}
\includegraphics[height=3in,width=5in]{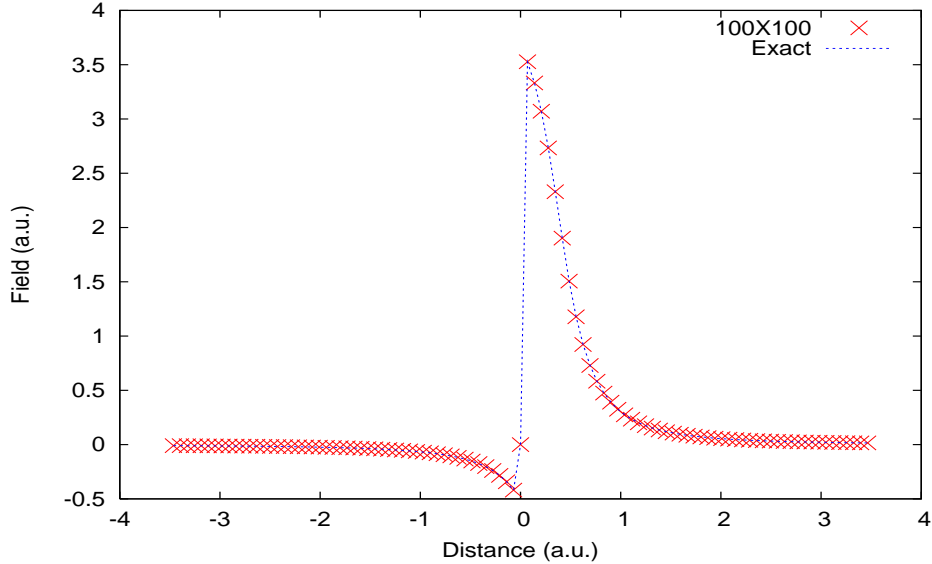}
\caption{\label{fig:TriDiag1FY} Comparison of flux (in the Y direction) as
computed by ISLES and numerical quadrature along a diagonal line}
\end{center}
\end{figure}

\begin{figure}[hbt]
\centering
\subfigure[Potential surface]{\label{fig:SurfTriElem}\includegraphics[width=0.45\textwidth]{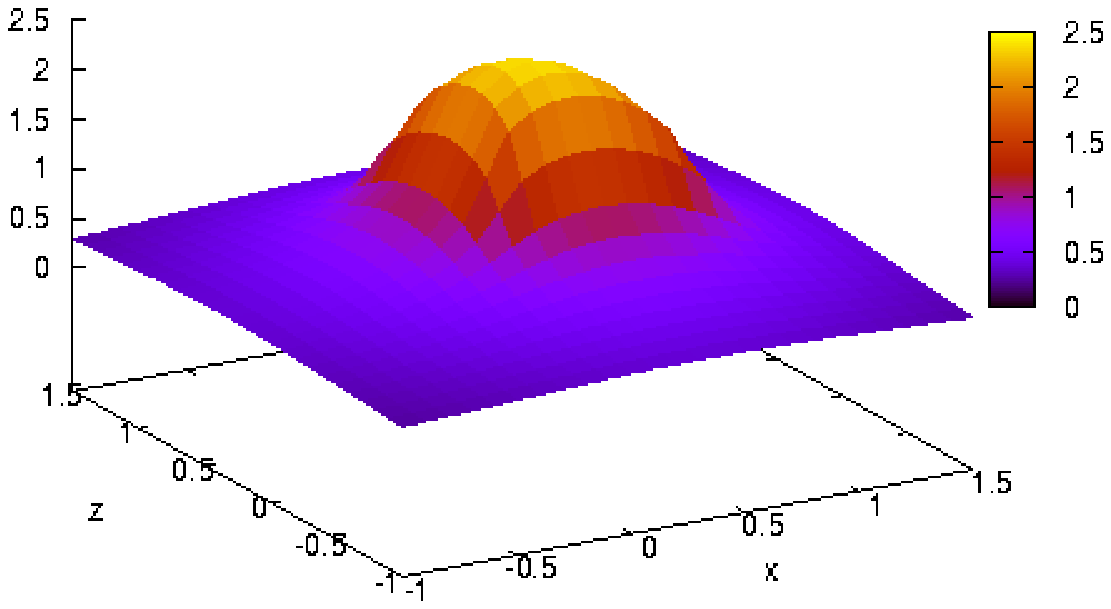}}
\subfigure[Flux surface]{\label{fig:TriFY_XY0}\includegraphics[width=0.45\textwidth]{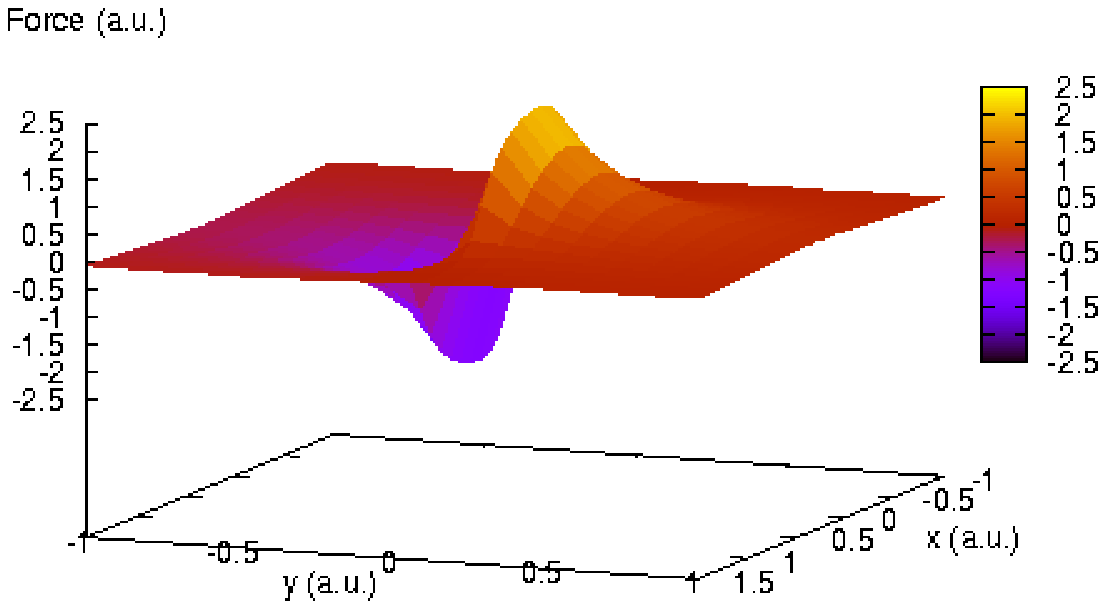}}
\caption{(a) Potential surface due to a triangular source distribution on the
element plane,
(b) Flux (in the Y direction) surface due to a triangular source distribution
on the XY plane at Z=0}
\label{fig:SurfacePlots}
\end{figure}

The expressions for the rectangular element have been validated in detail in
\cite{EABE2006}. Here, we present the results for triangular elements in fair
detail. In Fig.\ref{fig:CntrTriElem}, we have presented a comparison of
potentials evaluated for a unit triangular element by using the exact
expressions, as well as by using numerical quadrature of high accuracy. The two
results are found to compare very well throughout. Please note that contours
have been obtained on the plane of the element, and thus, represents a rather
critical situation. Similarly, Fig.\ref{fig:TriDiag1FY}
shows a comparison between the results obtained using closed-form expressions
for flux and those obtained using numerical quadrature. The flux considered here
is in the $Y$ direction and is along a line beginning from $(-2,-2,-2)$ and
ending at $(2,2,2)$. The comparison shows the commendable accuracy expected from
closed form expressions.
In Figs.\ref{fig:SurfTriElem} and \ref{fig:TriFY_XY0}, the surface plots of
potential on the element plane ($XZ$ plane) and $Y$-flux on the $XY$ plane
have been presented from which the expected significant increase
in potential and sharp change in the flux value on the element is observed.
Thus, by using a small
fraction of computational resources in comparison to those consumed in numerical
quadratures, ISLES can compute the exact value of potential and flux
for singularities distributed on triangular elements.

\subsubsection{Near-field performance}
\begin{figure}[hbt]
\begin{center}
\includegraphics[height=3in,width=5in]{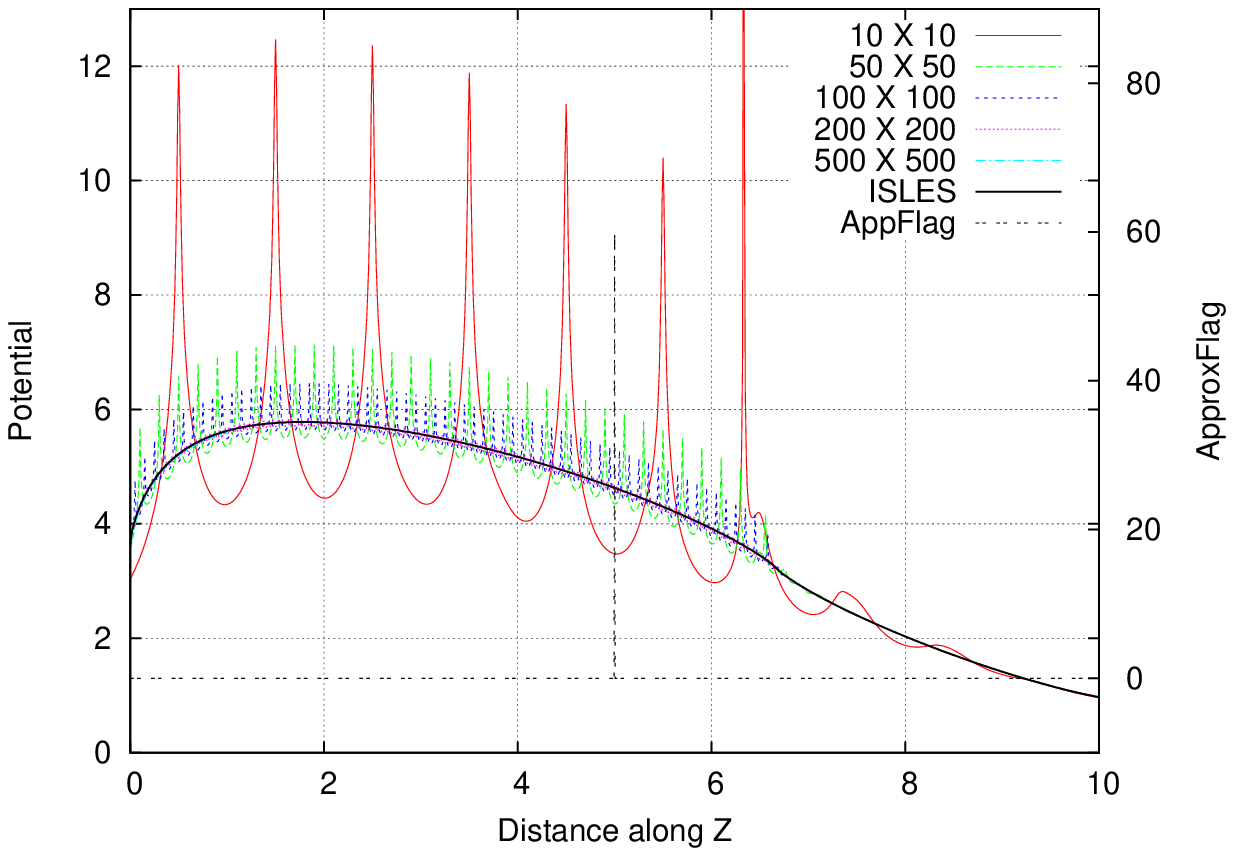}
\caption{\label{fig:TCentroidZVsPot} Variation of potential along a centroidal
line on the XZ plane parallel to the Z axis for a triangular element: comparison
among values obtained using the exact expressions and numerical quadratures}
\end{center}
\end{figure}

\begin{figure}[hbt]
\begin{center}
\includegraphics[height=3in,width=5in]{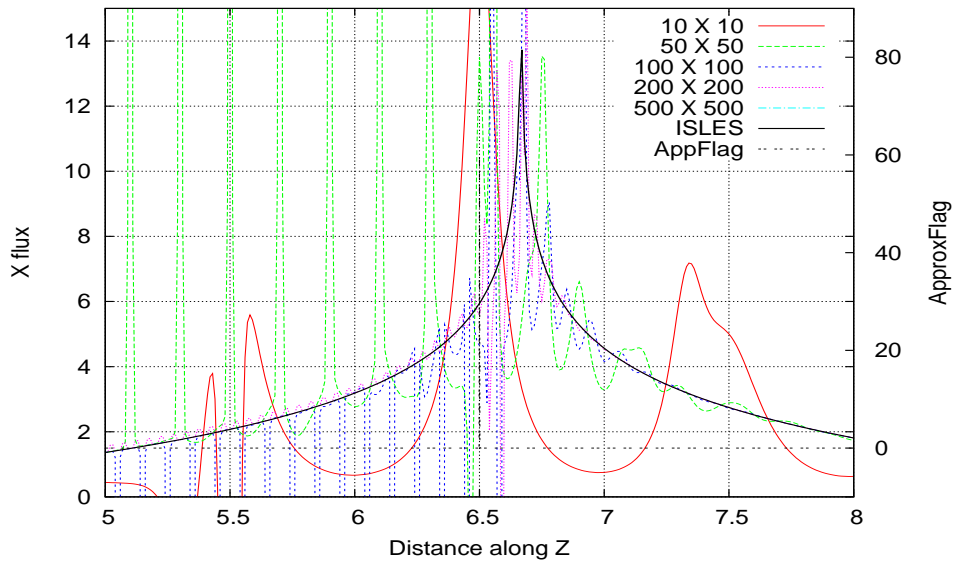}
\caption{\label{fig:TCentroidZVsFx} Variation of flux in the X direction along a
line on the XZ plane parallel to the Z axis for a triangular element: comparison
among values obtained using the exact expressions and numerical quadratures}
\end{center}
\end{figure}

\begin{figure}[hbt]
\begin{center}
\includegraphics[height=3in,width=5in]{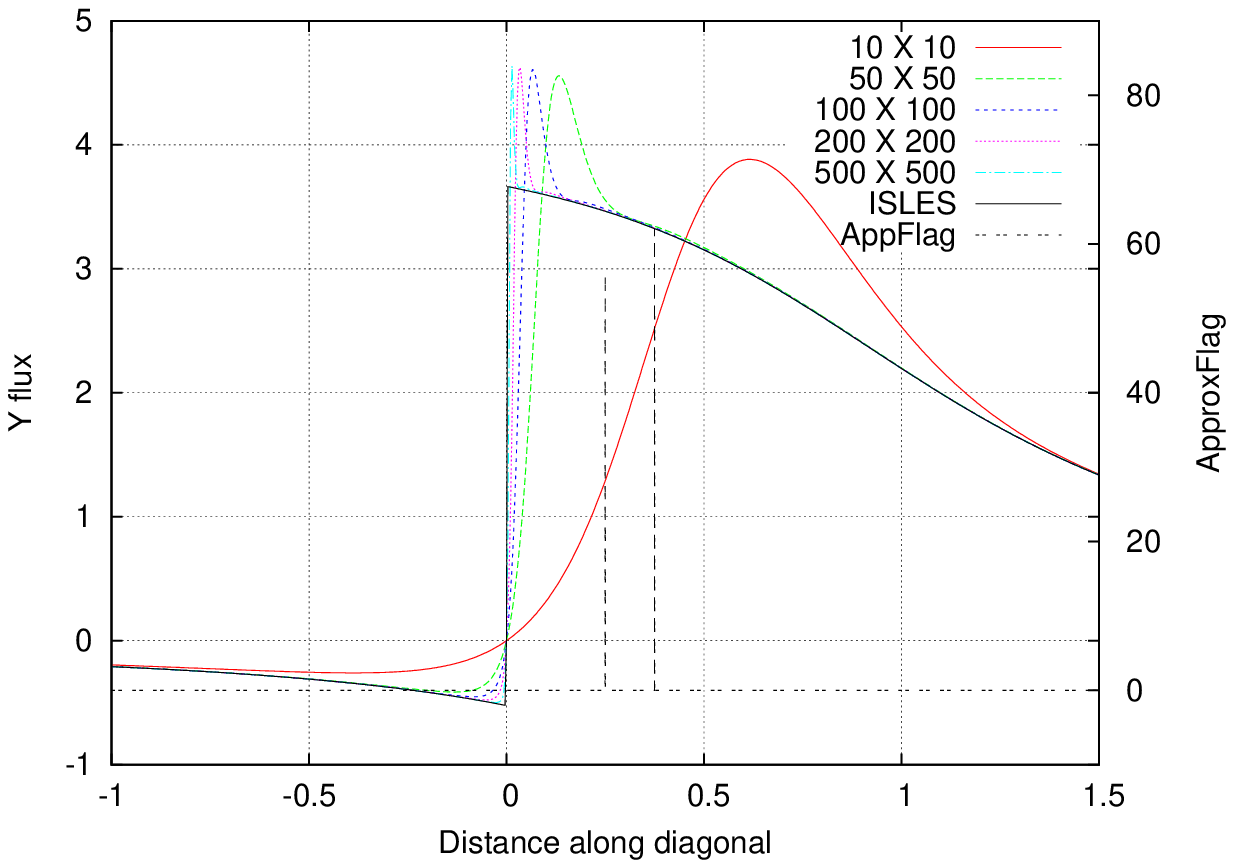}
\caption{\label{fig:TDiag1VsFy} Comparison of flux (in the Y direction) along a
diagonal line piercing the triangular element at the centroid: comparison
among values obtained using the exact expressions and numerical quadratures}
\end{center}
\end{figure}

In order to emphasize the accuracy of ISLES, we have considered the following
severe situations in the near-field region in which it is observed that the
quadratures can match the
accuracy of ISLES only when a high degree of discretization is used. Please
note that in these cases, the value of $z_M$ has been considered to be 10. In
Fig.\ref{fig:TCentroidZVsPot} we have presented the variation of potential
along a line on the element surface running parallel to the Z-axis of the
triangular element
(see Fig.\ref{fig:TriElemGeom}) and going through the centroid of the element.
It is observed that results obtained using even a $100 \times 100$ quadrature is
quite unacceptable. In fact, by zooming on to the image, it can be found that
only the maximum discretization yields results that match closely to the
exact solution. It may be noted here that the potential is a relatively easier
property to compute. The difficulty of achieving accurate flux estimates is
illustrated in the two following figures. The variation of flux in the
$X$-direction along the same line as used in Fig.\ref{fig:TCentroidZVsPot} has
been presented in Fig.\ref{fig:TCentroidZVsFx}. Similarly, variation of $Y$-flux
along a diagonal line (beginning at (-10,-10,-10) and ending at (10,10,10)
and piercing the element at the centroid) has been presented in
Fig.\ref{fig:TDiag1VsFy}. From these figures we see that the flux values
obtained using the quadrature are always inaccurate even if the discretization
is as high as $100 \times 100$.
We also observe that the estimates are locally inaccurate
despite the use of very high amount of discretization ($200 \times 200$ or
$500 \times 500$). Specifically, in the latter figure, even the highest
discretization can not match the exact values at the peak, while in the former
only the highest one can correctly emulate the sharp change in the flux value.
It is also heartening to note that the values from the quadrature using higher
amount of discretization consistently converge towards the ISLES values.

\subsubsection{Far field performance}
\label{sec:FarField}
It is expected that beyond a certain distance, the effect of the singularity
distribution can be considered to be the same as that of a centroidally
concentrated singularity or a simple quadrature. The optimized amount of
discretization to be used for the quadrature can be determined from a study of
the speed of execution of each of the functions in the library and has been
presented separately in a following sub-section. If we plan to replace the
exact expressions by quadratures (in order to reduce the computational expenses,
presumably) beyond a certain given distance, the quadrature should
necessarily be efficient enough to justify the replacement. While standard but
more elaborate algorithms similar to the fast multipole method (FMM)
\cite{Greengard87} along with the GMRES \cite{Saad86} matrix solver can lead to
further of computational efficiency, the simple approach as outlined
above can help in reducing a fair amount of computational effort. In the
following, we present the results of numerical experiments that help us in
determining the far-field performance of the exact expressions and quadratures
of various degrees that, in turn, help us in choosing the more efficient
approach for a desired level of accuracy.

\begin{figure}[hbt]
\begin{center}
\includegraphics[height=3in,width=5in]{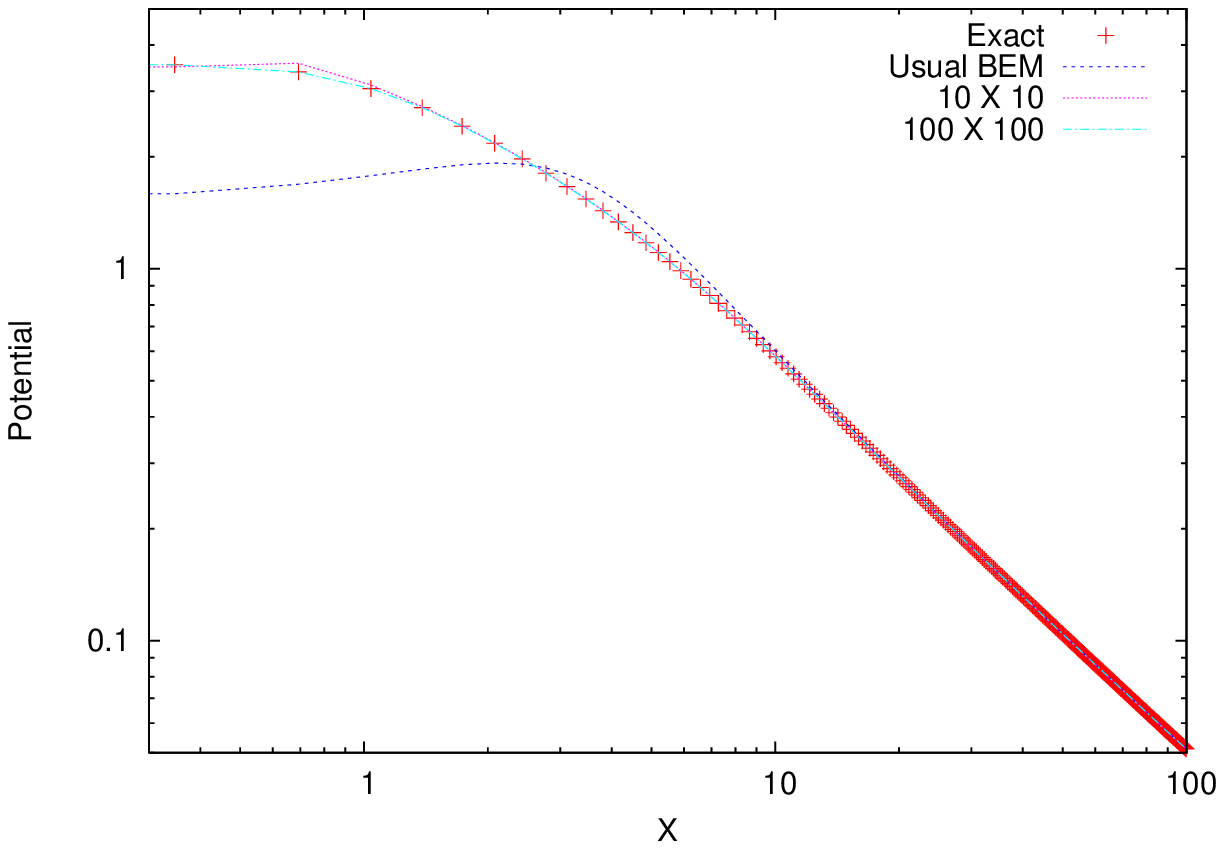}
\caption{\label{fig:DiagPFFarField} Potential along a diagonal through the
triangular element computed using exact, $100 \times 100$,  $10 \times 10$ and
usual BEM approach}
\end{center}
\end{figure}

\begin{figure}[hbt]
\begin{center}
\includegraphics[height=3in,width=5in]{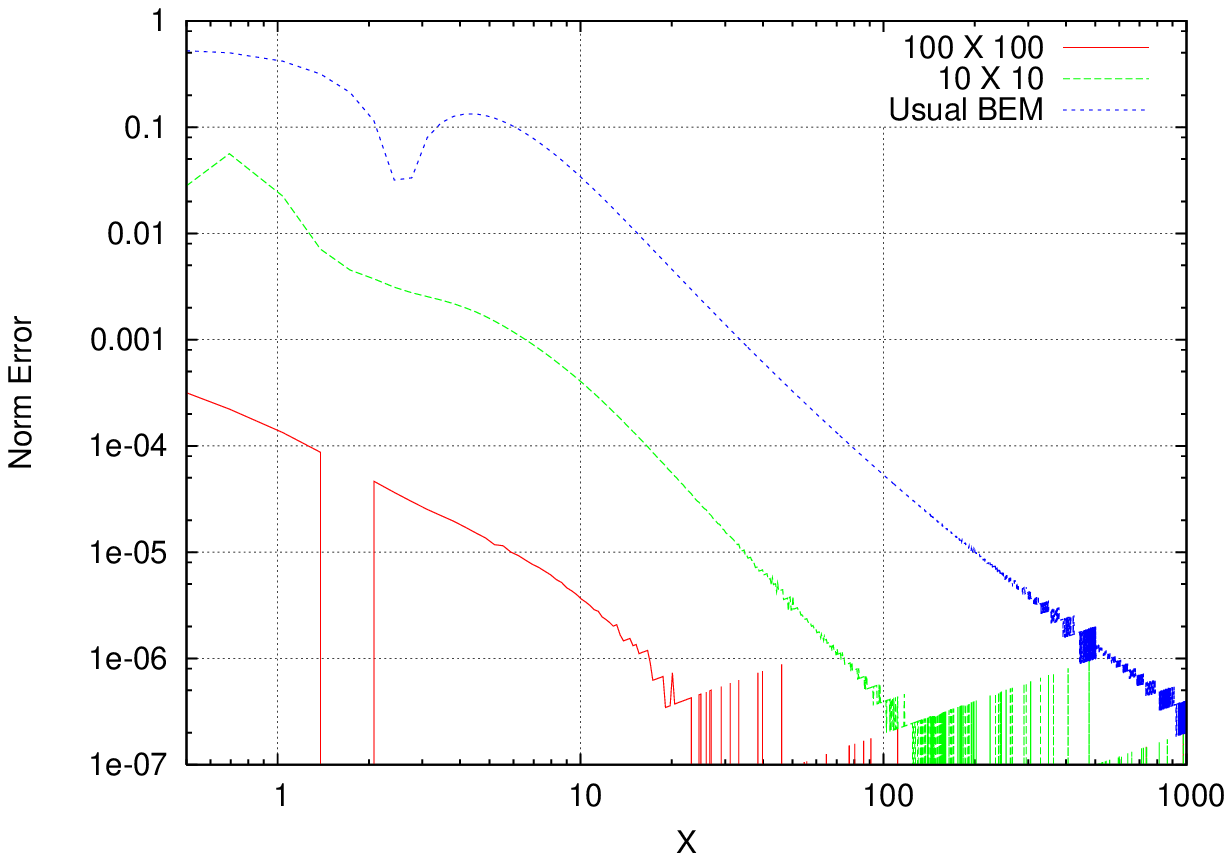}
\caption{\label{fig:ErrorPot} Error along a diagonal through the triangular
element computed using $100 \times 100$, $10 \times 10$ and usual BEM approach}
\end{center}
\end{figure}

In Fig.\ref{fig:DiagPFFarField} we have presented potential values obtained
using the exact approach, $100 \times 100$, $10 \times 10$ and no
discretization, i.e., the
usual BEM approximation while using the zeroth order piecewise uniform charge
density assumption. The potentials are computed along a diagonal line running
from (-1000, -1000, -1000) to (1000, 1000, 1000) which pierces a triangular
element of $z_M = 10$. It can be seen that results obtained using the usual BEM
approach yields inaccurate results as we move closer than distances of 10 units,
while the $10 \times 10$ discretization yields acceptable results up to a
distance
of 1.0 unit. In order to visualize the errors incurred due to the use of
quadratures, we have plotted Fig.\ref{fig:ErrorPot} where the errors
incurred (normalized with respect to the exact value) have been plotted. From
this figure we can conclude that for the
given diagonal line, the error due to the usual BEM approximation falls below
1\% if the distance is larger than 20 units while for the simple $10 \times 10$
discretization, it is 2 units. It may be mentioned here that along the axes the
error turns out to be significantly more \cite{EABE2006} and the limits
need to
be effectively doubled to achieve the accuracy for all cases possible. Thus,
for achieving 1\% accuracy, the usual BEM is satisfactory only if the distance
of the influenced point is five times the longer side of an element. Please note
here that the error drops to 1 out of $10^6$ as the distance becomes fifty times
the longer side. Besides proving that the exact expressions work equally
well in the near-field as well as the far-field, this fact justifies the usual
BEM approach for much of the computational domain leading to substantial savings
in computational expenses.

\subsubsection{Comparison with multipole expressions}

We also compared the value of potential with estimates for the barycenter
and other field point values as obtained from \cite{Lazic2006,Lazic2008}, the
expression having been slightly modified.
At the barycenter, the expression given is exact, whereas, the multipole
(monopole+quadrupole) expression is expected to be valid at any arbitrary
location.
The accuracy depends on the distance from the element as discussed in the
papers, an approximate rule being that the accuracy is of the order of $10^{-4}$
when the distance of the field point from the barycenter of the triangle is more
than $5.5$ times the longer side of the right triangle.
For points less distant than the mentioned value, the triangle needs to be
further segmented.
Consulting Fig. \ref{fig:ErrorPot} we may note that along a diagonally
intersecting line, the usual BEM achieves this accuracy only when the distance
of a field point is 8 times the larger side.
The $10 \times 10$ discretization is as good for a field point that is distant
twice the longer side.

In Table \ref{Table:CompareWithMultiPole}, we show the values estimated at the
barycenter by ISLES, analytic \cite{Lazic2006} and numerical quadrature of
different discretizations.
Triangular elements of different sizes have been used keeping
the x-side always of unit length.
It is interesting to note that the right triangle for which two perpendicular
sides are of equal length, it is most difficult to obtain the precise value
of potential using quadrature.
In fact, with a discretization of $2000 \times 2000$, we obtained the
value of 2.407462, while with $5000 \times 5000$ we obtained 2.407323.

\begin{table*}[hbt]
\centering
\caption{\label{Table:CompareWithMultiPole}Comparison of estimated potential by
ISLES, analytic and various quadratures}
\begin{tabular}{ l  c  c  c  c  c }
\hline
$z_M$ & ISLES & Analytic \cite{Lazic2006} & $10 \times 10$ & $100 \times 100$ & $500 \times 500$ \\
\hline
0.1    & 0.545069 & 0.545069 & 0.5410382 & 0.5450810 & 0.5450695 \\
1.0    & 2.407320 & 2.407320 & 2.460945   & 2.411947   & 2.408161 \\
10.0  & 5.450690 & 5.450690 & 5.257222   & 5.450847   & 5.450696 \\
\hline
\end{tabular}
\end{table*}

\begin{figure}[hbt]
\begin{center}
\includegraphics[height=3in,width=5in]{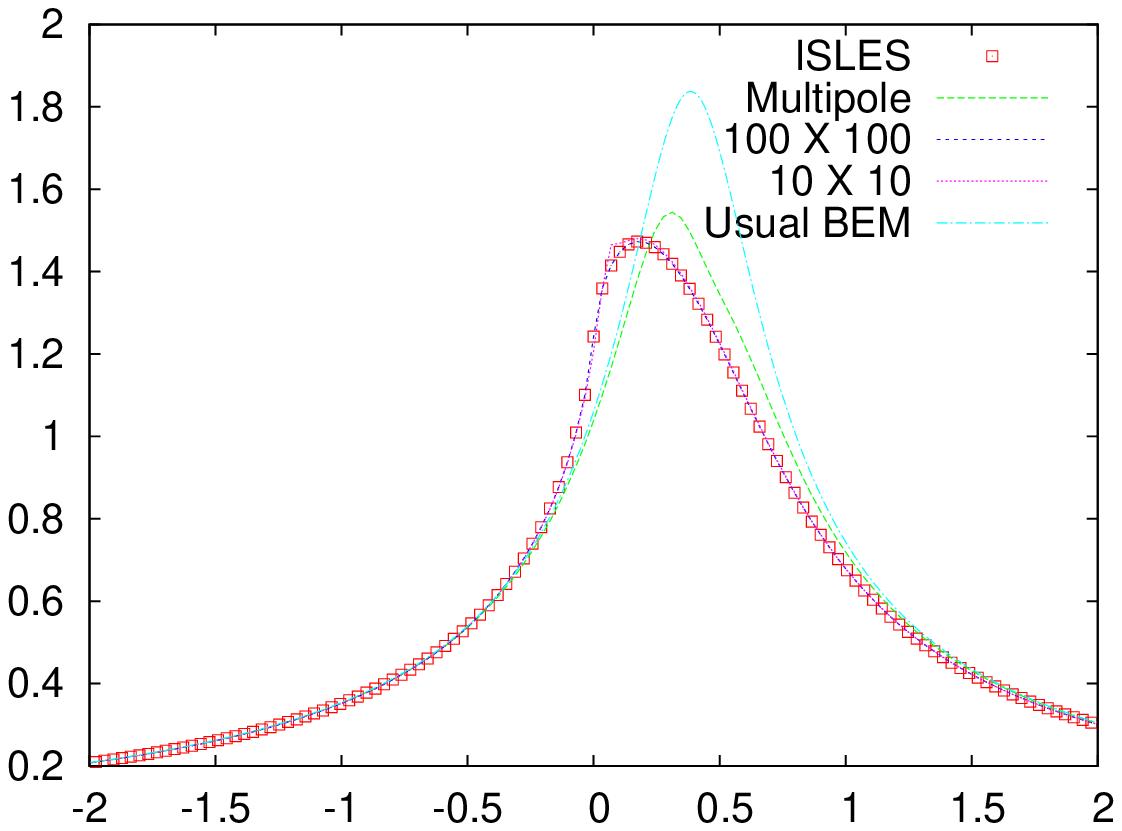}
\caption{\label{Fig:MultiPoleDiag1} Potential along a diagonal through the
triangular element computed using ISLES, multipole expansion,
$100 \times 100$,  $100 \times 10$ and usual BEM approach}
\end{center}
\end{figure}

\begin{figure}[hbt]
\begin{center}
\includegraphics[height=3in,width=5in]{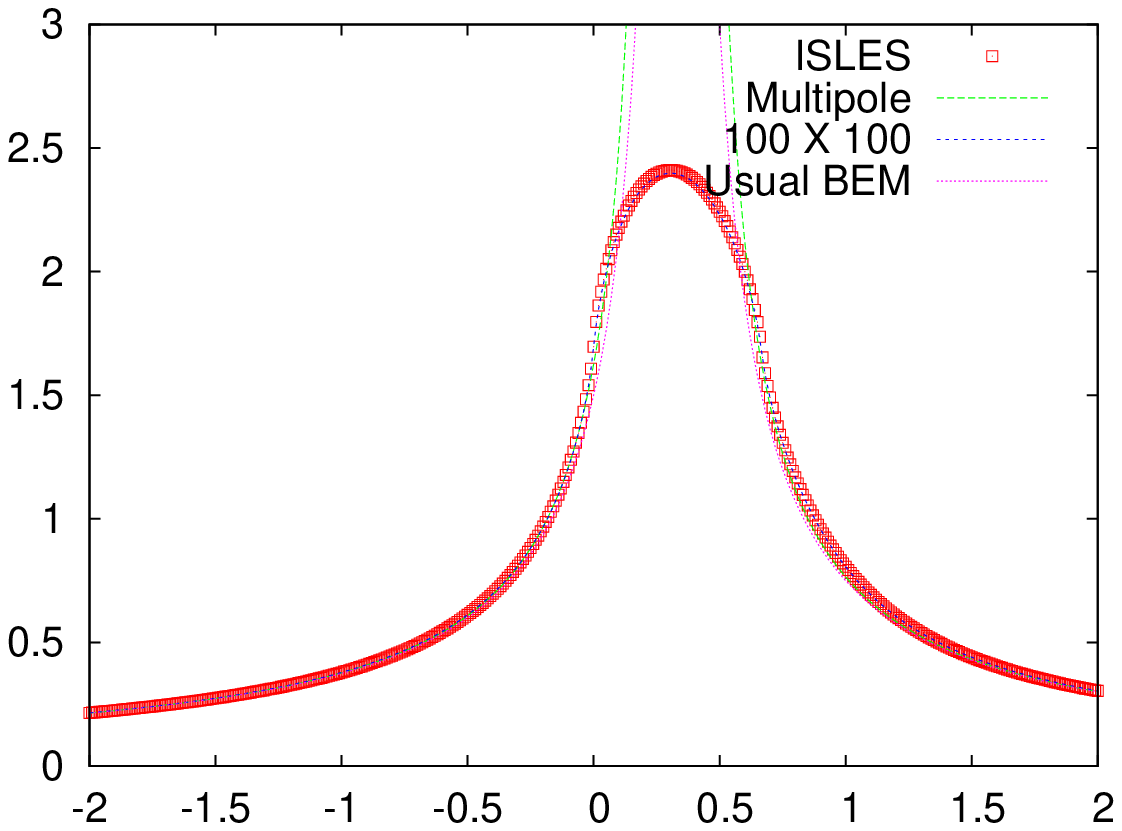}
\caption{\label{Fig:MultiPoleParallelX} Potential along a line parallel to X through
the centroid of a triangular element computed using ISLES, multipole expansion,
$100 \times 100$, $10 \times 10$ and usual BEM approach}
\end{center}
\end{figure}

In Fig.\ref{Fig:MultiPoleDiag1}, results from the above multipole expansion are
compared with those obtained using ISLES and numerical quadrature with different
levels of discretization along a line passing from (-10,-10,-10) to (10,10,10), the
range being reduced to a distance of -2 to +2 for ease of viewing.
Similar comparison has been carried  along a line parallel to the X-axis passing through
the barycenter of the element and in the same plane as the element in
Fig.\ref{Fig:MultiPoleParallelX}.
As indicated correctly in \cite{Lazic2006,Lazic2008}, the multipole expansion works
fine at distances far enough.
At close distances, only the ISLES results are acceptable.
Other options exist in terms of further discretization.
In that sense, both multipole expansion and numerical quadrature are likely
to work but it may be difficult to decide on the required level of discretization
\textit{a priori}.
Moreover, any advantage in terms of computational expenses may be lost due to the
necessity of increased discretization.

\subsubsection{Speed of execution}
The time taken to compute the potential and flux is an important parameter
related to the overall computational efficiency of the codes.
This is true despite the fact that, in a typical simulation, the time taken to solve
the system of algebraic equation is far greater than the time taken to build the
influence coefficient matrix and post-processing.
Moreover, the amount of time taken to solve the system of equations tend to
increase at a greater rate than the time taken to complete the other two.
Thus, in fact, evaluation of longer expressions in the pre- and post-processing
phase should hardly influence the efficiency of a BEM solver adversely.
Moreover, due to the enhanced accuracy of the proposed foundation
expressions, it will be possible to use significantly less number of elements
to represent a given device, ultimately leading to much faster execution
for a required level of accuracy.
It should be mentioned here that the time taken in each of these steps can
vary to a significant amount depending on the algorithm of the solver.
In the present case, the system of equations has been solved using lower upper
decomposition using the well known Crout's partial pivoting.
Although this method is known to be very rugged and accurate, it is not efficient
as far as number of arithmetic operations, and thus time, is concerned.
It is also possible to reduce the time taken to pre-process (generation of mesh and
creation of influence matrices), solve the system of algebraic equations and
that for post-process (computation of potential, flux at required
locations, estimation of other properties such as capacitance, force) by
adopting faster algorithms, including those involving parallelization. 

In order to optimize the time taken to generate the influence coefficient matrix
and that to carry out the post-processing, we carried out a small numerical
study to determine the amount of time taken to complete the various functions
being used in ISLES, especially those being used to evaluate the exact
expressions and those being used to carry out the quadratures.
The results of the study (which was carried out using the linux system command
\textit{gprof}) has been presented in the following
Table \ref{Table:SpeedTable}.

\begin{table*}[hbt]
\centering
\caption{\label{Table:SpeedTable}Time taken to evaluate exact expression of
ISLES, Usual BEM and
various quadratures}
\begin{tabular}{ l  c  c  c  c  c }
\hline
Method & ISLES       & Usual BEM & $10 \times 10$ & $100 \times 100$ & $500 \times 500$ \\
\hline
Time for  & 0.6 $\mu s$ & 25 $ns$   & 2 $\mu s$      & 400 $\mu s$      & 10 $ms$ \\
rectangular &                 &                 &                       &                           & \\
element   &                    &                 &                       &                           & \\
Time for  & 0.8 $\mu s$ & 25 $ns$   & 2 $\mu s$      & 400 $\mu s$      & 10 $ms$ \\
two triangular &             &                 &                       &                           & \\
elements &                    &                 &                       &                           & \\
\hline
\end{tabular}
\end{table*}

Please note that the numbers presented in this table are representative and
are likely to have statistical fluctuations.
However, despite the fluctuations,
it may be safely concluded that a quadrature having only $10 \times 10$
discretization is already consuming time that is comparable to that needed for
exact evaluation.
Thus, the exact expressions, despite their complexity, are extremely
efficient in the near-field which can be considered at least as large as 0.5
times the larger side of a triangular element (please refer to
Fig.\ref{fig:ErrorPot}).
In making this statement, we have assumed that the
required accuracy for generating the influence coefficient matrix and subsequent
potential and flux calculations is only 1\%.
This may not be acceptable at all under
many practical circumstances, in which case the near-field would imply a larger
volume.

Some of the advantages of using the ISLES library, based on the presented foundation
expressions, are mentioned below:
\begin{itemize}
\item{For a given level of discretization, the estimates are more accurate,}
\item{Effective efficiency of the solver improves, as a result,}
\item{Large variation of length-scales, aspect ratios can be tackled,}
\item{Thinness of members or nearness of surfaces does not pose any problem},
\item{Curvature has no detrimental effect on the solution,}
\item{The boundary condition can be satisfied anywhere on the elements,
i.e., points other than the centroidal points can be easily used, if necessary
(for a corner problem, may be),}
\item{The same formulation, library and solver is expected to work in majority
of physical situations. As a result, the necessity for specialized formulations
of BEM and associated complications can be greatly minimized.}
\end{itemize}

\subsection{Electrostatics of two- and three-dimensional bodies having corners
and edges}

\subsubsection{Square plate and Cube}
The capacitance value estimated by the present method has been compared with
very accurate results available in the literature (using BEM and other methods).
The results obtained using the neBEM solver is found to be among the most
accurate ones available till date as shown in Table \ref{table:CapCompSqr}.
Please note that we have neither invoked symmetry nor used extrapolation
techniques to arrive at our result presented in the table.
\begin{table*}[hbt]
\centering
\caption{\label{table:CapCompSqr}Comparison of capacitance values}
\begin{tabular}{ l  l  c  c }
\hline
Reference                    & Method               & Plate (pF) / 4 $\pi \epsilon_0$ & Cube (pF) / 4 $\pi \epsilon_0$ \\
\hline
\cite{Cavendish1879}and \cite{Maxwell1892} & SCM & 0.3638          &  \\
\cite{Reitan57}           & SCM                    & 0.362                                 & 0.6555 \\
\cite{Solomon}           & SCM                    & 0.367                                 & \\
\cite{Goto92} & Refined SCM & $0.3667894 \pm 1.1 \times 10^{-6}$& $0.6606747 \pm 5 \times 10^{-7}$  \\
                        & and Extrapolation           &                                           & \\
\cite{Read97} & Refined BEM & $0.3667874 \pm 1 \times 10^{-7}$   & $0.6606785 \pm 6 \times 10^{-7}$ \\
                        & and Extrapolation           &                                           &  \\
\cite{Mansfield2001} & Numerical Path    & 0.36684                             & 0.66069 \\
                        & Integration                      &                                           & \\
\cite{Wintle2004} & Random Walk           & $0.36 \pm 0.01$                & $0.6606 \pm 1 \times 10^{-4}$ \\
\cite{Mascagni2004} & Random Walk       &                                            & $0.6606780 \pm 2.7 \times 10^{-7}$ \\
\cite{Ong2005} & Singular element         &                                            & 0.6606749 \\
\cite{Read2004} & Refined BEM & $0.3667896 \pm 8 \times 10^{-7}$ & $0.6606767 \pm 4 \times 10^{-6}$ \\
                        & and Extrapolation           &                                           &                              \\
\cite{Lazic2008} & Robin Hood                 &                                            & $0.6606786 \pm 8 \times 10^{-8}$ \\
                            & and Extrapolation       &                                            & \\
This work               & neBEM                      &  0.3667524                         &  0.6606746 \\
\hline
\end{tabular}
\end{table*}

In Table \ref{Table:CubePotComparison}, we compare potentials at the center
and along an edge of the unit cube as obtained using neBEM with those from
\cite{Greenfield2004} in which the authors use analytical techniques to determine
the order of singularity at the singular regions of a cube.
From this table, we find that it has been possible for \cite{Greenfield2004}
to maintain accuracy of $10^{-6}$ in 1 for the potential values at the cube
center, $10^{-3}$ to $10^{-2}$ in 1 along an edge and $10^{-2}$ in 1 along
edge but close to a vertex of the cube.
For similar locations, results using neBEM indicate that an accuracy of $10^{-6}$
is maintained at the cube center, $10^{-3}$ at the edge and $10^{-2}$ on the
edge but close to the vertex.
Thus, the proposed approach has been able to achieve accuracy that is comparable
to those achieved by \cite{Greenfield2004} that uses the Fichera's theorem to
ensure proper variation  of singularities near edges and vertices.
Interestingly, at critical locations near the cube vertex, results from neBEM are found
to be better than \cite{Greenfield2004} by a significant amount.
We have been able to maintain this accuracy of $10^{-2}$ as close as upto
$1 \mu m$ to the cube vertex which is extremely encouraging.
Unfortunately, we have not been able to compare our results with other numerical
results at distances less than a mm from the vertex.
It is encouraging to note, however, that while for \cite{Greenfield2004}, the error
is $4.8 \times 10^{-3}$ when the evaluation point is $1 mm$ away from the
vertex, neBEM commits an error of a similar amount only when the point is
$10 \mu m$ away from the vertex.
The error for neBEM becomes larger by a small amount ($0.4 \times 10^{-3}$) only
when the evaluation point is as close as a micron to the vertex.
\begin{table*}[hbt]
\centering
\caption{\label{Table:CubePotComparison}Comparison of potential at the center and
along an edge of a unit cube}
\begin{tabular}{ c  c  c  c  c  c  c  c }
\hline
X        &Y	   & Z     & Exact & \cite{Greenfield2004} & Error in \cite{Greenfield2004} & neBEM & Error in neBEM \\
\hline
0        & 0    & 0     & 1        & 0.999990  & $-1.0 \times 10^{-5}$ & 1.000001   & $1.0 \times 10^{-6}$ \\
0.4     & 0.5 & 0.5  & 1        & 0.9996      & $-4.0 \times 10^{-4}$ & 0.9994362 & $-5.638 \times 10^{-4}$ \\
0.45   & 0.5 & 0.5  & 1        & 0.99986   & $-1.4 \times 10^{-4}$  & 0.9995018 & $-4.982 \times 10^{-4}$ \\
0.49   & 0.5 & 0.5  & 1        & 1.0013     & $1.3 \times 10^{-3}$   & 0.9991151 & $-8.849 \times 10^{-4}$ \\
0.499 & 0.5 & 0.5  & 1        & 1.0048     & $4.8 \times 10^{-3}$   & 0.9987600 & $-1.24 \times 10^{-3}$ \\
0.4999 & 0.5 & 0.5  & 1      & -               & -                                     & 0.9974398 & $-2.56 \times 10^{-3}$ \\
0.49999 & 0.5 & 0.5  & 1    & -               & -                                     & 0.9951335 & $-4.8 \times 10^{-3}$ \\
0.499999 & 0.5 & 0.5  & 1  & -               & -                                     & 0.9945964 & $-5.4 \times 10^{-3}$ \\
\hline
\end{tabular}
\end{table*}

\begin{figure}[hbt]
\begin{center}
\includegraphics[height=3in,width=5in]{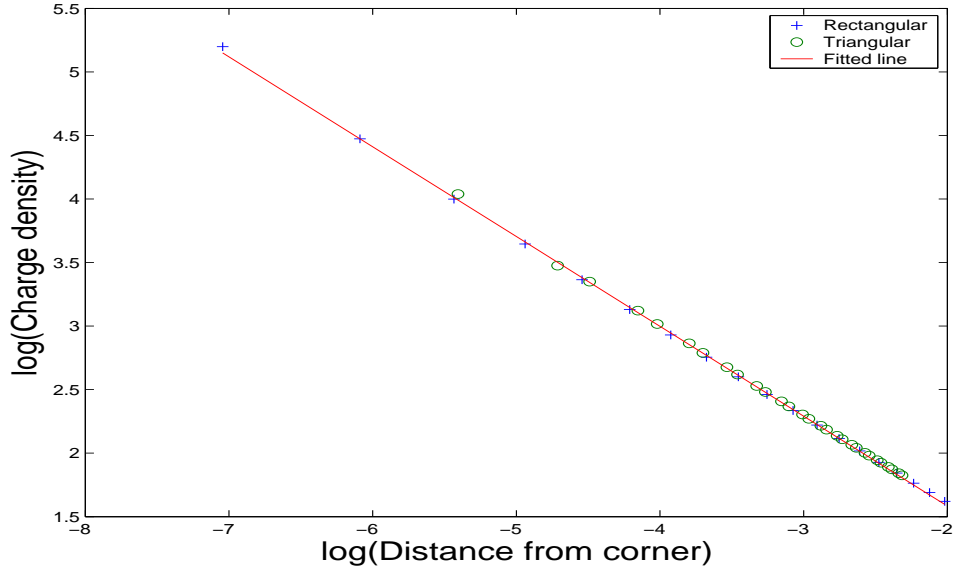}
\caption{\label{fig:FitCornerChDen} Variation of charge density with increasing
distance from the corner of the unit square plate and a least-square fitted
straight line: slope of the fitted line is 0.7068}
\end{center}
\end{figure}

Next, we consider the problem of determining charge density distribution at
corners and edges of the above geometries.
Problems of this nature are considered to be challenging for any numerical tool and
especially so for the BEM approach.
In Table \ref{table:SingOrderSqrCube}, we have presented the estimates of the order
of singularity at the vertex or the edge as done by methods as diverse as singular
perturbation \cite{Morrison76}, BEM \cite{Read2004}, last-passage
and walk on spheres \cite{Given2003,Mascagni2004}, application of Fichera's
theorem \cite{Greenfield2004}, singular element approach \cite{Ong2005} and
the presented approach.
From the table, it is clear that there is good agreement among all the methods.
As in \cite{Read2004} properties on the element next to a corner or edge 
has been ignored while carrying out the least-square fits.
Points were included in the fit as long as the maximum mismatch between the
fitted line and the computed value was less than typically 1\%, which also
allows us to include points as long as they fall closely on a straight line.
Following this approach, we could use values of singularities even upto 0.15 (for
a plate or cube of unit side) from the relevant edge or corner
while fitting the lines (see Fig.\ref{fig:FitCornerChDen}, as an example where we used
both triangular and rectangular elements for discretizing a square plate and obtained
a singularity index of 0.7057 and 0.7068, respectively).
These can be compared to the facts that in \cite{Read2004}, two points next to
a corner was excluded, as well as all points at distances beyond 0.05 from a corner.
However, the exact value of the index is, to a certain extent, dependent on the
discretization and the details of the least square fitting procedure.
Thus, it may not be very prudent to attach great significance to the obtained values
except noting that they agree with each other and also agree with the theoretical
estimates, wherever available.
Thus, it may be difficult to accurately ascertain the singularity index at an arbitrary
corner or edge.
This difficulty can lead to problems for methods that depend on beforehand knowledge
of the order of singularity.

\begin{table*}[hbt]
\centering
\caption{\label{table:SingOrderSqrCube}Comparison of estimation of order of singularity}
\begin{tabular}{ l  l  c  c  c  c }
\hline
Reference & Method 								& Plate corner & Plate edge & Cube vertex &Cube edge \\
\hline
                 &                                             &                     &                  &                       & \\
\cite{Jackson} & Analytic                        &                     & 0.5            &                       & 0.333 \\
\cite{Morrison76} & Numerical shooting & 0.7034         &                  &                       & \\
\cite{Schimtz93}  & BEM 	                     &                     &                  & 0.5468           & \\
\cite{Given2003} & Walk on spheres 	& 0.7034          &                  & 0.5381 /         & \\
                            &                                  &                     &                   & 0.5458           & \\
\cite{Read2004} & Surface Charge 		& 0.704            &                  &  0.540            & \\
\cite{Mascagni2004} & Surface Charge &                     &                   & 0.558            & 0.333 \\
\cite{Ong2005} & Singular element       &                     &                   & 0.5475          & \\
\cite{Wintle2004} & Random Walk         &                     &                   &                     & \\
\cite{Greenfield2004} & Fichera's theorem & 0.7015    &                   & 0.5454         & \\
\cite{Hwang2005} & Walk on planes      & 0.7034         &                   & 0.5457         & \\
This work               & neBEM                     & 0.7068      & 0.4994       & 0.5539           & 0.332 \\
\hline
\end{tabular}
\end{table*}

\begin{figure}[hbt]
\begin{center}
\includegraphics[height=3in,width=5in]{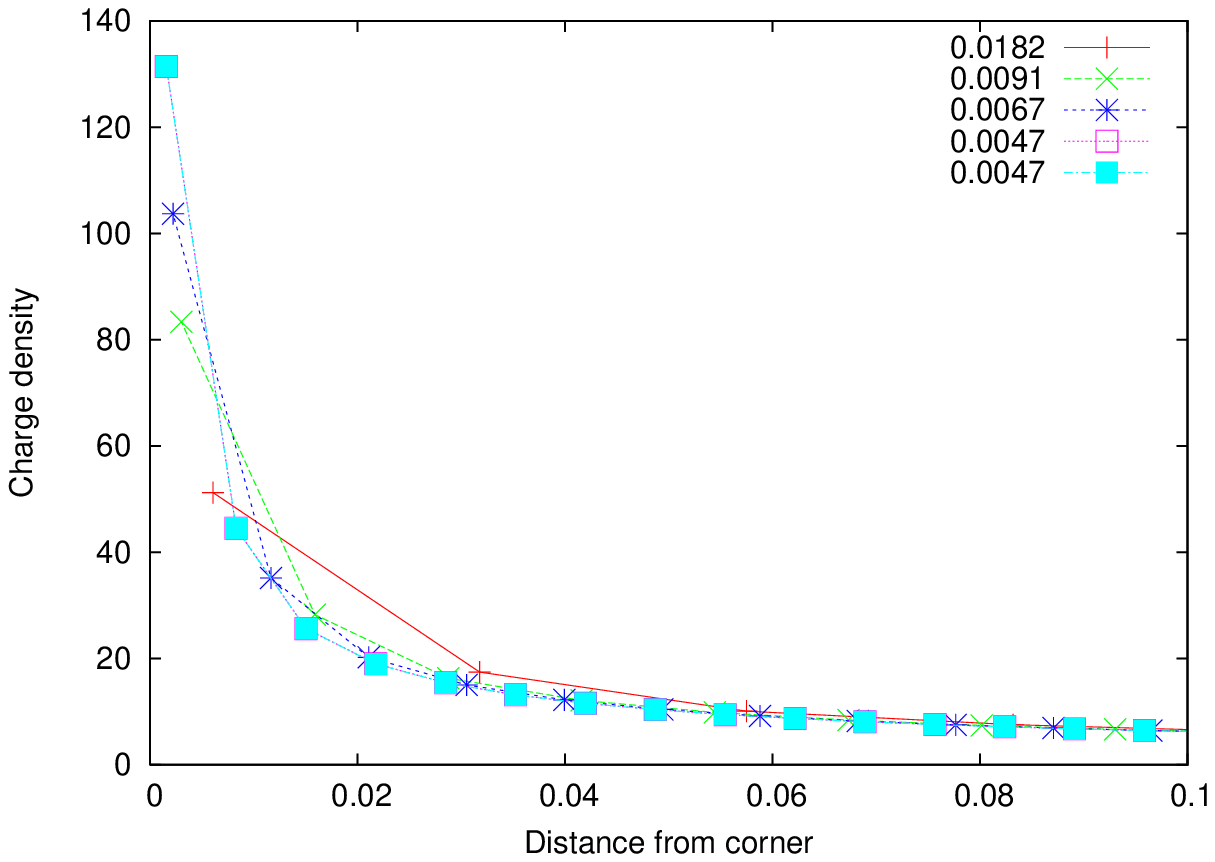}
\caption{\label{fig:CompareCornerChDen} Corner charge density estimated by 
\textit{neBEM} using various sizes of triangular elements}
\end{center}
\end{figure}

In the following study, we have presented estimates of charge density very close
to the flat plate corner as obtained using neBEM.
This has been carried out to investigate the objection raised against the
BEM in \cite{Wintle2004} where the author states that severe oscillations in the charge
densities are expected close to the corner and edge of plates because
the BEM cannot correctly model the edge / corner of physical devices.
Please note that for this study, the boundary
conditions have been satisfied at the centroids of each element although the
neBEM has the capability of satisfying boundary conditions at locations other
than the centroid.
In
Fig.\ref{fig:CompareCornerChDen}, charge densities very close to the corner
of the flat plate estimated by neBEM using various amounts of discretization
have been presented.
It can be seen that each curve follows the same general
trend, does not suffer from any oscillation and is  found to be converging to a
single curve.
This is true despite the fact that there has been almost an order of
magnitude variation in the element lengths.
Thus, we can safely conclude that the estimates obtained using neBEM do
not suffer from the numerical instabilities mentioned in \cite{Wintle2004}.

\begin{figure}[hbt]
\begin{center}
\includegraphics[height=3in,width=5in]{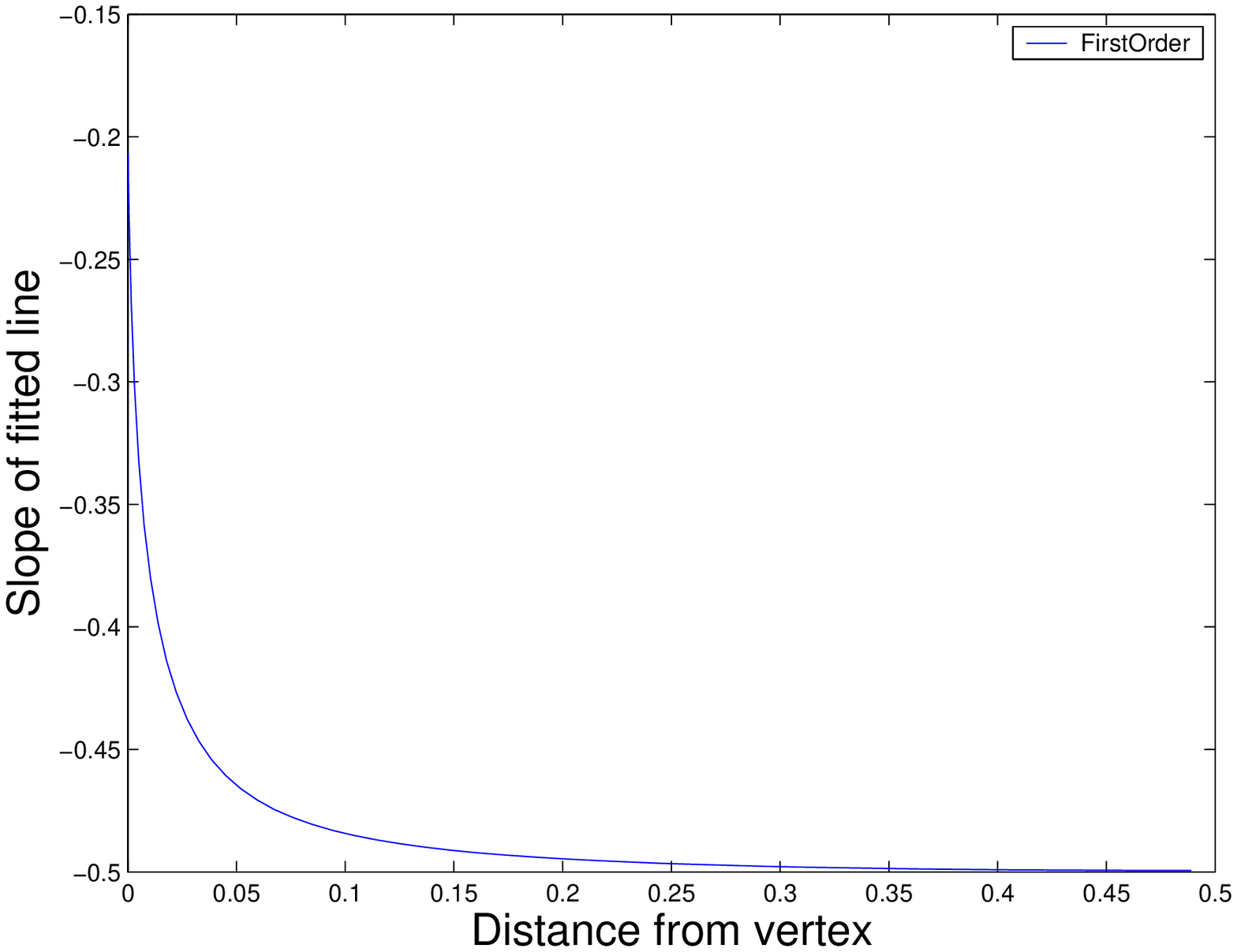}
\caption{\label{fig:VariationOfSingularitySqrPlate} Variation of the slope of the
fitted lines along an edge of a square plate}
\end{center}
\end{figure}

In Fig.\ref{fig:FitCornerChDen}, we present a least-square fitted
straight line matching the charge density as obtained using the highest
discretization in this study.
Results using both triangular and rectangular elements are presented
and it is found that the slope of the fitted line is 0.7057 when the elements
are triangular, whereas, it is 0.7068 when we use rectangular elements.
Both the values compare very well with both old and recent estimates of the
order of singularity as shown in Table \ref{table:SingOrderSqrCube}.
In Fig.\ref{fig:VariationOfSingularitySqrPlate}, we have shown how
the slope of the fitted line changes along the edge of a square plate as we
move away from a corner of a square plate.
From the figure, it is apparent that the change in the singularity index along an edge
of a square plate, can be quite significant and only when we are in reality
close to the middle of the edge, the analytic value of 0.5 can be used with
confidence for the order of singularity.
This observation is significant especially for the singular element and singular
function methods where prior knowledge of this parameter plays a crucial
role in determining the accuracy of the solution.

\subsubsection{L-shaped plate and volume}
\begin{figure}[hbt]
\begin{center}
\includegraphics[height=3in,width=5in]{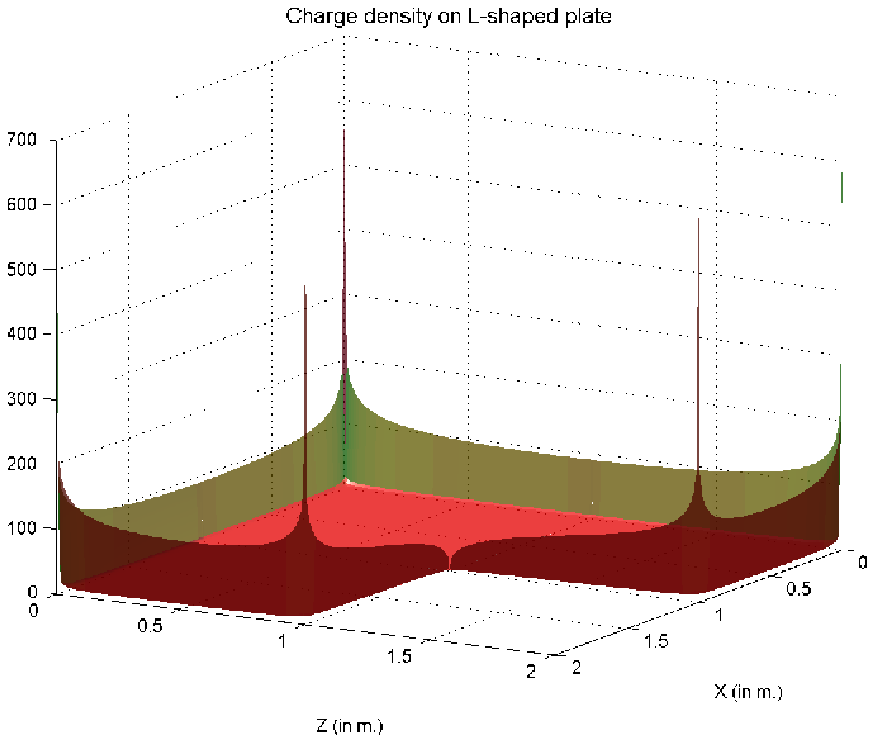}
\caption{\label{fig:ChDenSurfOnLShapedPlate} Charge density estimated by 
\textit{neBEM} on a L-shaped conducting plate raised to unit volt}
\end{center}
\end{figure}

Despite being geometries of generic nature, these L-shaped geometries have received
relatively less attention.
Here, we have estimated the capacitance and orders of
singularity at various important locations, e.g., inner and outer corners
and compared these values with available numerical results.
It should be mentioned here that in \cite{Ong2005,Su2002}, the L-shaped volume
has been described very nicely in relation to the varying nature of
singularities it contains.
Capacitance of the L-shaped volume conductor has turned out to be 112.1497pF.
The estimate matches extremely well with the value of 112.15pF that has been
used as a reference value in \cite{Ong2005}.
In Table \ref{table:SingOrderLShaped} we present the numerical values of the
order
of singularities at inner corners of the L-shaped plate and volume conductor.
While the former matches well with \cite{Morrison76}, the latter is not quite close
to the estimate in \cite{Ong2005}, the difference being close to 20\%.
\begin{table*}[hbt]
\centering
\caption{\label{table:SingOrderLShaped}Comparison of estimation of order of singularity}
\begin{tabular}{ l  l  c  c }
\hline
Reference & Method 								 & Plate inner & L volume inner \\
                 &                                             & corner        &  corner       \\
\hline
\cite{Morrison76} & Numerical shooting & 0.1854       &  \\
\cite{Ong2005} & Singular element       &                    & 0.1104 \\
This work               & neBEM                    & 0.1840        & 0.0896 \\
\hline
\end{tabular}
\end{table*}
Finally, in Fig.\ref{fig:ChDenSurfOnLShapedPlate}, we show how the magnitude
of charge density changes on a L-shaped plate.
The remarkable difference between external and internal corners in terms of
charge density concentration is very clearly observed in this figure.

\subsubsection{Edge problem having analytical solutions in 2D}
While none of the above problems have analytic solutions, a closely related
problem has well-known analytic solution in the two-dimensional case
\cite{Jackson}.
In fact, while discussing the earlier problems, other
workers and we have quite often referred to this solution obtained in standard
textbooks as an exercise in the method of separation of variables.

We have considered a three-dimensional equivalent of the geometry as presented
in \cite{Jackson} in which two conducting planes intersect each other at an angle
$\beta$.
The planes are assumed to be held at a given potential.
In order to specify the boundary conditions conveniently, a circular cylinder is 
also included that just encloses the two intersecting plane, has its center at the
intersection point and is held at zero potential \cite{Integrated}.
The general solution in the polar coordinate
system ($\rho$, $\phi$) for the potential ($\Phi$) close to
the origin in this problem has been shown to be
\begin{equation}
\label{eq:EdgePotential}
\Phi(\rho, \phi) = V\,
+\, {\sum_{m=1}^{\infty}}\, a_m\, \rho^{m \pi / \beta}\,sin(m \pi \phi/\beta)
\end{equation}
where the coefficients $a_m$ depend on the potential remote from the corner at
$\rho = 0$ and $V$ represents the boundary condition for $\Phi$ for all $\rho
\geq 0$ when $\phi = 0$ and $\phi = \beta$. In the present case where a circular
cylinder just encloses the two plates, the problem of finding out $a_m$
reduces to a basic fourier series problem with a well known solution
\begin{equation}
\label{eq:Coeff}
a_m\, =\, \frac{4}{m \pi}\,\,\, for\, odd\, m
\end{equation}
It may be noted here that the series in (\ref{eq:EdgePotential}) involve
positive powers of $\rho^{\pi/\beta}$, and, thus, close to the origin (i.e., for
small $\rho$), only the first term in the series will be important. The electric
field components ($E_{\rho}, E_{\phi}$) are
\begin{equation}
\label{eq:Efield_rho}
E_{\rho}(\rho,\phi) =
-\frac{\pi}{\beta}
{\sum_{m=1}^{\infty}}\, a_m\, \rho^{(m \pi/\beta) - 1} sin(m \pi \phi / \beta)
\end{equation}
\begin{equation}
\label{eq:Efield_phi}
E_{\phi}(\rho,\phi) =
-\frac{\pi}{\beta}
{\sum_{m=1}^{\infty}}\, a_m\, \rho^{(m \pi/\beta) - 1} cos(m \pi \phi / \beta)
\end{equation}
The surface charge densities ($\sigma$) at $\phi = 0$ and $\phi = \beta$ are equal and
are approximately
\begin{equation}
\label{eq:ChDen}
\sigma(\rho)\, =\, \frac{E_{\phi}(\rho, 0)}{4\pi}
\simeq -\frac{a_1}{4 \beta}\, \rho^{(\pi/\beta) - 1}
\end{equation}
Thus, the field components and the charge density near $\rho = 0$ both vary with
distance as $\rho^{(\pi/\beta) - 1}$ and this fact is expected to be reflected
in a correct numerical solution as well.

While the above theoretical solution is a two-dimensional one, we have used the
neBEM solver to compute a three-dimensional version of the above problem.
In order to reproduce the two-dimensional behavior at the mid-plane, we have
made the axial length of the system sufficiently long, viz., 10 times the
radius of the cylinder.
The radius of the cylinder has been fixed at one meter,
while the length of the intersecting flat plates has been made a mm shorter
than the radius to avoid the absurd situation of having two values of the
voltage at the same spatial point.

The cylinder has been discretized uniformly in the angular and axial directions.
The flat plates have also been uniformly discretized in the axial direction.
In the radial direction, however, the flat plate elements have been made
successively smaller towards the edges using a simple algebraic equation in
order to take care of the fact that the surface charge density increases
considerably near the edges.
From Tables \ref{table:Edge360}, \ref{table:Edge270} and \ref{table:OtherEdges},
we can compare the accuracy of neBEM results with other analytical and
numerical results.
The two ends of the range of angles, 360 and 90, represent particularly
difficult situations.
The former is difficult due to the very large concentration of charge density
close to a sharp edge and the resulting large electric field.
The latter is difficult due to the fact that to truly simulate a null point in a
concave corner, extremely precise estimates are necessary to ensure
cancellation of electric field.
Throughout the range, the neBEM results are found to be very accurate
except at the location that is just $1 \mu m$ away from the edge.
Even at this location, for all the convex edges, the results are reasonable
and surely comparable to the only other numerical result available.
The neBEM estimates are unacceptable, however, at locations less than
tens of microns away from the null corner of the concave corner.
It may be noted however, that the problem is not at all inherent to the
formulation and is clearly related to the size of elements used in the
vicinity of the corner.
Here, on our desktop PC with 2GB RAM, we could use a spatial resolution
of around a micron close to the corner despite using a large profiling
factor.
And this was at the steep cost of having elements with extremely large
aspect ratios ($1:10^6$).
We believe that these are the factors that have resulted in the inaccuracy
of the presented results when the locations considered were a micron
away from the edge.

\begin{table*}[hbt]
\centering
\caption{\label{table:Edge360}Electric field close to a 360deg edge}
\begin{tabular}{ l  l  c  c  c  c }
\hline
Distance                & Analytical   & ELECTRO   & Error (\%)    & neBEM       & Error (\%) \\
\hline
0.8                         & 0.3954180 & 0.3954213 & 0.00059       & 0.3950786 & -0.086 \\
0.1                         & 1.830153   & 1.830155   & 0.00010       &  1.830110  & -0.002 \\
0.01                       & 6.303166   & 6.303172   & 0.000094     &  6.305784  & -0.041 \\
0.001                     & 20.11157   & 20.11122   & 0.0018         &  20.11963  & -0.04 \\
0.0001                   & 63.65561   & 63.64274   & 0.020           &  63.64780  & -0.012 \\
0.00001                 & 201.3148   & 200.88       & 0.22             &  200.5488  & -0.3 \\
0.000001               & 636.6191   & 621.25       & 2.4               &  621.6034  & -2.36 \\
\hline
\end{tabular}
\end{table*}

\begin{table*}[hbt]
\centering
\caption{\label{table:Edge270}Electric field close to a 270deg edge}
\begin{tabular}{ l  l  c  c  c  c }
\hline
Distance                & Analytical       & ELECTRO   & Error (\%)    & neBEM       & Error (\%)   \\
\hline
0.8                         & 0.5246997      & 0.524710  & 0.0019         & 0.5241510 & -0.105 \\
0.1                         & 1.747623        & 1.747621  & 0.00014       &  1.747953  & -0.018 \\
0.01                       & 3.931433        & 3.931284  & 0.0038         &  3.933242  & -0.046 \\
0.001                     & 8.487415        & 8.4854      & 0.023           &  8.491335  & -0.046 \\
0.0001                   & 18.28732        & 18.202      & 0.46             &  18.29270  & -0.029 \\
0.00001                 & 39.39902        & 35.80        & 9.1               &  39.30955  & -0.227 \\
0.000001               & 84.88264        & 57.10        & 32.7             &  80.74309  & -4.877 \\
\hline
\end{tabular}
\end{table*}

\begin{table*}[hb]
\centering
\caption{\label{table:OtherEdges}Electric field close to 90deg and 225deg edges}
\begin{tabular}{ l  c  c  c  c  c  c  c }
\hline
Distance          & 0.8            & 0.1                & 0.01               & 0.001            & 0.0001            & 0.00001        & 0.000001 \\
\hline
                        &                  &                      &                       & 90 degrees   &                        &                       & \\
\hline
Analytic    & 1.445221   & 2.546224e-1 & 2.546479e-2 & 2.546479e-3 &  2.546479e-4  & 2.546479e-5 & 2.546479e-6 \\
neBEM       & 1.444098   & 2.547710e-1 & 2.548018e-2 & 2.547723e-3 &  2.778723e-4  & 3.760117e-3 & 1.518850 \\ 
Error (\%)  & -0.001123  & 0.058361 & 0.060436           & 0.048846       &  9.120200       & usable(?)      & inaccurate \\ 
\hline
                        &                  &                      &                       & 225 degrees   &                        &                       & \\
\hline
Analytic     & 0.6266090 & 1.574802      & 2.556973      & 4.055022      &  6.426876        & 10.18592      & 16.14359 \\
neBEM       & 0.6259308 & 1.575165      & 2.557946      & 4.056545      &  6.428439        & 10.16933      & 15.63912 \\ 
Error (\%)  & -0.108233  & -0.023050     & -0.038052     & -0.037558    &  -0.024313       & -0.162871     & -3.124893 \\ 
\hline
\end{tabular}
\end{table*}

To end this section, we present one contour plot of electric field for a
convex corner.
From the Fig.\ref{fig:ConvexEdgeContour}, it is evident that the intensity of
the field increases only very close to the convex edge.
Please note that the dimensions in the figure are mentioned in microns.
\begin{figure}[hbt]
\begin{center}
\includegraphics[height=3in,width=5in]{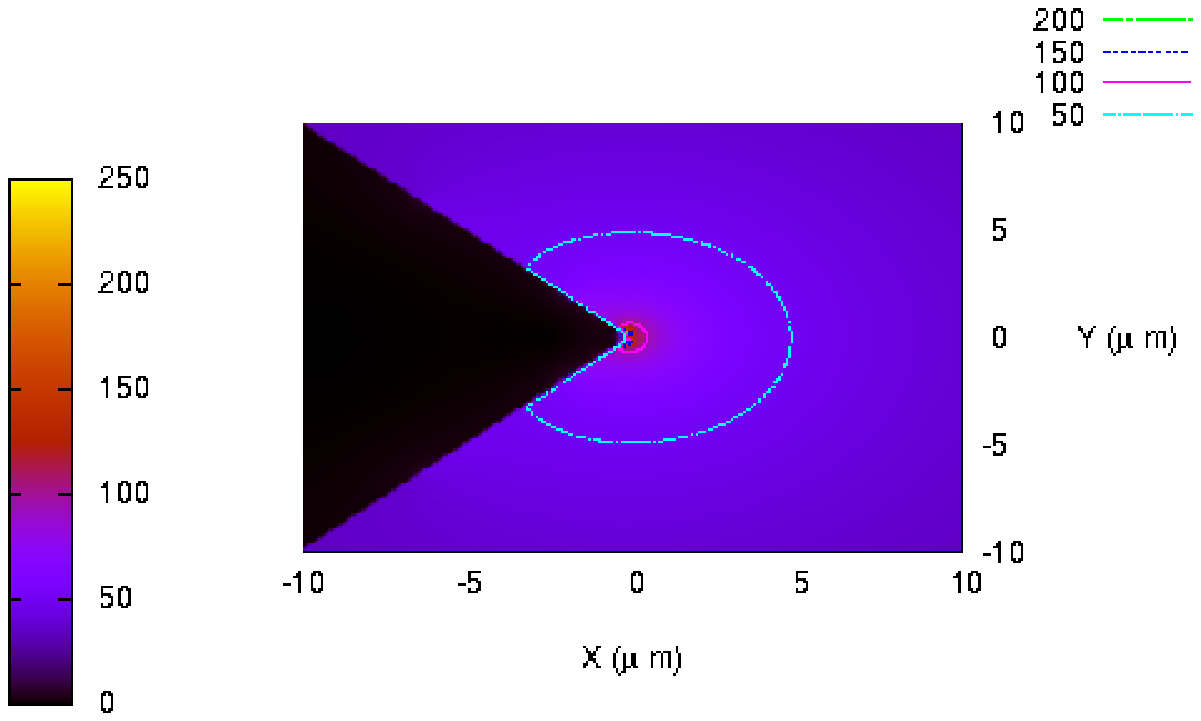}
\caption{\label{fig:ConveXEdgeContour} Electric field distribution very close
to a convex edge.}
\end{center}
\end{figure}

\section{Conclusion}
An efficient and robust library, ISLES, for solving potential problems in a large
variety of science and engineering problems has been presented.
Exact closed-form expressions used to develop ISLES have been validated throughout
the physical domain (including the critical near-field region) by comparing
these results with results obtained using numerical quadrature of high
accuracy and with those obtained using quadrupole expressions.
Algorithmic aspects of this development have also been touched upon.
The neBEM solver that uses foundation expressions being evaluated by
ISLES, has been used to solve several corner and edge electrostatic problems.
Several classic benchmark problems such as those related to unit square
plate and unit cube have been solved to very high precision.
Charge density values at critical geometric locations like corners
have been found to be numerically stable and physically acceptable.
Values of singularity indices at different corners and edges have been estimated
and compared with other analytic and numerical estimates.
The agreement among the different approaches has been found to be
quite acceptable.
It has been observed that the variation of this index along the edge of a
conducting body is non-negligible implying caution necessary for methods
that need prior knowledge of these indices to solve a given problem.
Finally, using the same solver, a three-dimensional equivalent of an edge
problem has been solved for which analytic solution exists in two-dimensions.
Detail comparison with this problem and those stated earlier have led us to
believe that the neBEM approach is a precise, flexible and robust solver
that works over a very wide range of problems.
Several advantages over usual BEM solvers and other specialized BEM solvers
have been briefly mentioned.
Some of the criticisms leveled against the BEM approach in general have been
addressed in this work, while we expect to solve some of the remaining in
future communications.
Work is also under way to make the code more efficient through
the implementation of faster algorithms and parallelization.

\vspace{0.25in}
\textbf{Acknowledgements}\\
We would like to thank Professor Bikas Sinha, Director, SINP and Professor
Sudeb Bhattacharya, Head, INO Section, SINP for their support and encouragement
during the course of this work.

%
%

\end{document}